\documentclass{aa}  
\usepackage[colorlinks=true, linkcolor=blue, citecolor=blue, urlcolor=blue]{hyperref}
\usepackage{graphicx}
\usepackage{tabularx}
\usepackage{breqn}
\usepackage{float}
\usepackage{txfonts}
\usepackage{natbib}

\begin{document} 

\title{Radio timing constraints on the orbital orientation and component masses of PSR J1455$-$3330}
\titlerunning{Mass measurements and 3D orbital geometry of PSR J1455$-$3330}

\author{D.S. Pillay\inst{1},
        V. Venkatraman Krishnan \inst{1}, David J. Champion \inst{1}, P. C. C. Freire \inst{1}, M. Kramer \inst{1,2}, L. Guillemot \inst{3,4}, M. Bailes \inst{5,6}, A. Corongiu \inst{7}, M. Geyer \inst{8}, J. Singha \inst{8}, R. M. Shannon\inst{5,6}, G. Theureau \inst{3,4}, I. Cognard \inst{3,4}, M. J. Keith \inst{2}, B. W. Stappers \inst{2}, J. Antoniadis \inst{9}, K. Liu \inst{10,1}, G. M. Shaifullah\inst{11,12,7}}
\authorrunning{D. S. Pillay}

\institute{Max-Planck-Institut für Radioastronomie, Auf dem Hügel 69, D-53121 Bonn, Germany \\  \email{dpillay@mpifr-bonn.mpg.de} 
\and  Jodrell Bank Centre for Astrophysics, University of Manchester, M13 9PL, UK 
\and LPC2E, OSUC, Univ Orleans, CNRS, CNES, Observatoire de Paris, F-45071 Orleans, France 
\and ORN, Observatoire de Paris, Universit\'e PSL, Univ Orl\'eans, CNRS, 18330 Nan\c{c}ay, France  
\and Centre for Astrophysics and Supercomputing, Swinburne University of Technology, Hawthorn, VIC 3122, Australia.
\and The Australian Research Council Centre of Excellence for Gravitational Wave Discovery (OzGrav)  
\and  INAF - Osservatorio Astronomico di Cagliari, Via della Scienza 5, I-09047 Selargius (CA), Italy  
\and  High Energy Physics, Cosmology \& Astrophysics Theory (HEPCAT) Group, Department of Mathematics and Applied Mathematics, University of Cape Town, Rondebosch 7701, South Africa 
\and  FORTH Institute of Astrophysics, N. Plastira 100, 70013, Heraklion Greece 
\and State Key Laboratory of Radio Astronomy and Technology, Shanghai Astronomical Observatory, CAS, 80 Nandan Road, Shanghai 200030, P. R. China
\and Dipartimento di Fisica ``G. Occhialini'', Universit\`{a} degli Studi di Milano-Bicocca, Piazza della Scienza 3, I-20126 Milano, Italy 
\and INFN, Sezione di Milano-Bicocca, Piazza della Scienza 3, 20126 Milano, Italy }

\date{12 January 2026}

\abstract{\rm PSR J1455$-$3330 is a $\sim$7.98 ms pulsar in a $\sim$76.17 day nearly circular orbit with a white dwarf companion. In this work, we combine the available Lovell,  Nan\c{c}ay decimetric Radio Telescope, Green Bank, and MeerKAT pulsar timing data spanning $\sim$ 30 years to measure the kinematic and relativistic effects of PSR J1455$-$3330 to constrain its 3D orbital geometry and component masses. We detect a relativistic Shapiro delay signal. We measure a significant orthometric amplitude $h_3 = 0.307^{+0.022}_{-0.026}$ $\mu$s and an orthometric ratio $\varsigma = 0.551^{+0.057}_{-0.054}$. We measure the change in projected semi-major axis $\dot{x} = -202.1^{+2.5}_{-2.7} \times10^{-16} \, \rm s\,s^{-1}$ with high significance, parallax, $\varpi$ = 1.11(6) mas, parallax derived distance 0.90(5) kpc, and a precise total proper motion magnitude of 12.432(2) mas yr$^{-1}$. A self-consistent analysis of all kinematic and relativistic effects, assuming general relativity, yields two solutions: (1) a pulsar mass $M_{\rm p} = 1.39^{+0.38}_{-0.18}\, \rm M_{\odot}$, a companion mass $M_{\rm c} = 0.293^{+0.056}_{-0.026}$ $\rm M_{\odot}$, an orbital inclination, $i = 63(2)^{\circ}$, and longitude of the ascending node, $\Omega = 212(12)^{\circ}$ or (2)  a pulsar mass $M_{\rm p} = 1.53^{+1.10}_{-0.22} \, \rm M_{\odot}$, a companion mass $M_{\rm c} = 0.309^{+0.163}_{-0.026}\, \rm M_{\odot}$, an orbital inclination, $i = 123(4)^{\circ}$, and longitude of the ascending node, $\Omega = 334(12)^{\circ}$. All uncertainties represent the 68.27$\%$ credibility region. These results strongly favour a helium-dominated white dwarf companion.}


\maketitle

\section{Introduction}
Rotating neutron stars that emit beams of electromagnetic radiation from their magnetic poles (i.e. pulsars) are observed as a steady, periodic sequence of pulses as the beam sweeps through our line-of-sight. This lighthouse-like rotation of a pulsar can be tracked with high precision using pulsar timing and can offer a wealth of information in the field of fundamental physics. 

Precise mass measurements of neutron stars (NSs) are important for studies regarding equations of state (EoS), ultra stripped supernovae explosions, and binary evolution theories. The current observable mass distribution ranges from $\sim 1.174(4)\, \rm M_{\odot}$ \citep[PSR J0453+1559c]{2015ApJ...812..143M} to $\sim 2.08(7)\, \rm M_{\odot}$ \citep[PSR J0740+6620]{2021ApJ...915L..12F}. The highest masses can be used to study EoS, the lowest masses can be used to study ultra-stripped supernovae explosions \citep{2015ApJ...812..143M} and everything in between is valuable for studies of binary evolution.

Millisecond pulsars (MSPs) are ``recycled'' neutron stars that have extremely stable spin periods ($P$ $\lesssim$ 100 ms and spin-down $\dot{P} \lesssim$ 10$^{-17}$ s $\rm{s^{-1}}$) whose pulse times-of-arrival (ToAs) can be measured with great precision. Around $80\%$ of MSPs are found in binary systems, with companions of varying mass, including main sequence stars, neutron stars or white dwarfs.

MSPs with helium white dwarf (HeWD) companions, in particular, form after Gyrs of accreting matter onto the neutron star from a low-mass star \citep{2023pbse.book.....T}. During this stage, the system is visible at X-ray wavelengths as a low-mass X-ray binary. The eccentricities of these binaries are well explained by the orbital period–eccentricity relation \citep{1992RSPTA.341...39P}, where most binary millisecond pulsars have very low orbital eccentricities. The tidal interaction with the red-giant companion circularises the orbits. After Roche-lobe overflow, the pulsar is spun up to millisecond periods, and the companion is left as a HeWD \citep{1982CSci...51.1096R,1982Natur.300..728A, 2023pbse.book.....T}. In such systems, the binary orbital period, $P_{\rm b}$, and mass of the HeWD, M$_{\rm WD}$, are correlated following the \citet{1999A&A...350..928T} relation.

In this work, we discuss our timing analysis of the MSP system PSR J1455$-$3330 (hereafter J1455$-$3330) using data from the Lovell telescope, Nan\c{c}ay decimetric Radio Telescope (NRT), Green Bank telescope and MeerKAT telescope. J1455$-$3330 was discovered in a Southern Hemisphere survey search for millisecond pulsars along the Galactic disk using the 64-m CSIRO Parkes Murriyang radio telescope \citep{1995ApJ...439..933L}.
The short spin period ($P \sim 7.98$ ms), and small spin down rate ($\dot{P} \sim 2.43\times 10 ^{-20}\, \rm s s^{-1}$) indicate that the pulsar is recycled. 
The small mass function  ($f = 0.00627\, \rm M_\odot$), under the assumption that the pulsar has a canonical mass of 1.4\,$\rm M_\odot$, \citep{1995ApJ...439..933L} results in a minimum companion mass of $M_{\rm c} = 0.27\, \rm M_\odot$. This minimum companion mass and the nearly circular orbit ($e \sim$ 0.00016) indicate that the companion is a white dwarf.

The \citet{2023A&A...678A..48E} detected a highly significant (33 $\sigma$, where 1 $\sigma$ is the 68.27\% credibility level) change in its projected semi-major axis ($\dot{x}$) and a moderate (2\,$\sigma$) $\dot{P}_{\rm b}$ value using the available Lovell and Nan\c{c}ay data. The detection of $\dot{x}$ provides us with an opportunity to constrain the 3D orbital geometry of the system as in PSR J1933$-$6211 \citep{2023A&A...674A.169G}, PSR J2222$-$0137 \citep{2021A&A...654A..16G} and PSR J0955$-$6150 \citep{2022A&A...665A..53S}.

The stable timing properties, previously detected PK parameters, and close proximity (DM = 13.56 cm$^{-3}$ pc) make J1455$-$3330 a favourable system for component mass measurements. J1455$-$3330 has been observed with MeerKAT from April 2019 to April 2024 under the MeerTime Large Survey Project \citep{2020PASA...37...28B} as part of two sub-projects: the relativistic binary timing programme \citep[RelBin]{2021MNRAS.504.2094K} and the pulsar timing array programme \citep[MPTA]{2022PASA...39...27S}. The main objectives of Relbin are to observe compact relativistic binary pulsar systems to measure potential relativistic effects, test theories of gravity, and measure NS masses. The PTA programme aims to search for low-frequency gravitational waves by regularly observing and timing many stable millisecond pulsars.

In this paper, we aim to place constraints on the component masses, using the available Lovell, Nan\c{c}ay, Green Bank and recently acquired MeerKAT data. We use 
the measured proper motion, change in the projected semi-major axis, and the Shapiro delay signal detected for the first time to constrain the system's component masses and orbital orientation.

The structure of this paper is as follows. In Sect. \ref{Section2} we discuss the observations and the preliminary data analysis to obtain the required times of arrivals. In Sect. \ref{Section3}, we describe the polarisation profile of the pulsar and the fit of its linear polarisation using the Rotating Vector Model \citep[RVM]{1969ApL.....3..225R}. In Sect. \ref{Section4} we describe the timing analyses, orbital model selection, and noise model selection. The timing results, measured post-keplerian parameters, component mass estimations, and orbital orientation constraints are presented in Sect. \ref{Section5}. In Sect. \ref{Section6}, we discuss the implications of the results, infer the formation and evolution of the system, and discuss the future prospects. 

\section{Observations, data acquisition and reduction}\label{Section2}
In this work, we use data from 4 different telescopes: the 76-m
Lovell Telescope in the
United Kingdom, the 94-m equivalent NRT in France, the 100-m Green Bank Telescope (GBT) in the United States, and the MeerKAT 64-dish
array in South Africa.

All radio observations and data reduction are detailed below and listed in Table \ref{Tab:1}. 

\begin{table*}[t]
\centering
\caption{\rm Observing systems and timing datasets of J1455$-$3330.}
	\label{Tab:1}
\begin{tabular}{l|ll|ll|llll|l} 
        \cline{1-10}
            Telescope & \multicolumn{2}{c}{JBO} & \multicolumn{2}{c}{Nan\c{c}ay} &  \multicolumn{4}{c}{GBT} & MeerKAT \\
        \cline{1-10}
        \cline{1-10}
            Receiver & \multicolumn{2}{c}{L-band} &  \multicolumn{2}{c}{L-band} &  \multicolumn{2}{c}{Rcvr_800} &\multicolumn{2}{c}{Rcvr_1_2} & L-band \\
            \cline{2-10} Backend&AFB&DFB&BON&NUPPI&GASP&GUPPI&GASP&GUPPI&PTUSE\\
            CF (MHz)&1402&1520&1400&1484&844&820.5&1410&1518&1283.58\\
            BW (MHz)&64&380&64/128&512&64&180&64&640&856\\
        \cline{1-10}
            Start MJD&48920&54849&53375&55819&53217& 
            55278 &53219&55429&58589\\
            End MJD &55114&60471&55812&60113&55578&
            58940&55245&58939&60411\\
            No. of ToAs &297&67&671&3617&479 
            &4843&564&4932&991\\
        \cline{1-10}
            Weighted rms residual ($\mu$s) &64.406 &24.187 & 5.282&2.648 & 4.835 &3.918 &5.679 &3.748 &1.895\\
            Reduced $\chi^2$ &1.01&0.96&0.98&0.99&0.94&
            0.93&1.02&1.04&0.98\\
            EFAC&1.07 &1.21&1.06&1.30&0.99&
            1.02&0.99&1.02&1.07\\
            log$_{10}$(EQUAD)&$-$4.38&$-$7.09&$-$9.83& $-$7.84& $-$5.89
            &$-$8.61& $-$5.89&$-$8.61&$-$6.3\\
        \cline{1-10}   
\end{tabular}
\tablefoot{The table shows the telescopes, receivers, backend specifications, associated centre observing frequency (CF), the effective observable bandwidth (BW), the number of frequency channels (Nchans), the time span of the dataset, the timespan in years, the number of ToAs, best-fit reduced $\chi^2$
derived from a subsequent TEMPO2 fit, and the derived white-noise parameters EFAC and EQUAD, which were determined with \texttt{temponest}.}
\end{table*}

\subsection{Lovell Telescope}
J1455$-$3330 was typically observed for 30 minutes per session, once every 10 days with a subintegration time of 10 s \citep{2004MNRAS.353.1311H}. The data used in this work were recorded between October 1992 and June 2024 with an analogue filterbank (AFB) until January 2009 and the digital filterbank (DFB) thereafter. The data recorded with the AFB have a central frequency of 1402 MHz and an observing bandwidth of 64 MHz. The DFB is a clone of the Parkes Digital FilterBank with a central observing frequency of 1400 MHz, and a bandwidth of 128 MHz split into 512 channels. From September 2009, the centre frequency was changed to 1520 MHz and the BW increased to 512 MHz (split into 1024 channels) of which $\sim$ 380 MHz was usable, depending on RFI conditions. 
There are no standard polarisation calibrations applied to the DFB data, but the power levels of both polarisations are manually and regularly adjusted via a set of attenuators. An on-site H-maser clock is used to estimate local time, which is corrected to UTC using recorded offsets between the maser and GPS satellites \citep{2016MNRAS.458.3341D}.

The data were first incoherently dedispersed to the DM of the pulsar and folded at the best-known period. These folded profiles were cleaned using the \texttt{PSRCHIVE paz} and \texttt{pazi} tools, and the ToAs were generated by cross-correlating an analytic noise-free template with the time-integrated, frequency-scrunched, total intensity profile. The template was generated using the \texttt{paas} tool to fit a set of von Mises functions to a profile created using \texttt{psradd} on the high signal-to-noise (S/N) ratio observations.

\subsection{Nan\c{c}ay decimetric Radio Telescope}
We selected NRT observations carried out with the Berkeley-Orl\'eans-Nan\c{c}ay (BON) backend from January 2005 until August 2011, and with the NUPPI backend (a version of the Green Bank Ultimate Processing Instrument, installed at the NRT) after it became the primary pulsar timing instrument in August 2011, until June 2023. BON observations were carried out at 1.4~GHz, over a bandwidth of 64~MHz until mid-2008, and then over 128~MHz. NUPPI observations were done at a central frequency of 1.484~GHz, and over a larger bandwidth of 512~MHz. BON and NUPPI observations were coherently dedispersed, and were typically $\sim$ 1 h long. 

Daily observations were fully scrunched in time. BON observations were scrunched in frequency in order to form two frequency sub-bands of 32~MHz (resp. 64~MHz) for the data taken before (resp. after) mid-2008. NUPPI data were split into four sub-bands of 128~MHz each. High S/N template profiles for the BON and NUPPI datasets were created by summing the eight observations with the highest S/N, and smoothing the results. ToAs were finally extracted from each observation using the Matrix Template Matching \citep[MTM,][]{2006ApJ...642.1004V} implemented in the \texttt{pat} routine of PSRCHIVE. Additional information about the NRT and about the preparation of NRT pulsar timing data can be found in \citet{2023A&A...678A..79G}.

\subsection{Green Bank Telescope}
Observations were taken at 1.4 GHz using the Rcvr\_1\_2 receiver, and 800 MHz using the Rcvr\_800 receiver with two different backends (GASP and GUPPI).
\newline
At 1.4 GHz, we use the data recorded with the GASP backend, with bandwidth $\sim$ 64 MHz from August 2004 until February 2010. Similarly, at 1518 MHz, we use the data recorded with the GUPPI backend with a bandwidth of 640 MHz from August 2010 until March 2020. 

At 800 MHz we use the GASP backend observations (at central frequency 844 MHz, with bandwidth $\sim$ 64 MHz) from July 2004 until January 2011 and the GUPPI backend observations (at central frequency 820.5 MHz, with bandwidth $\sim$ 180 MHz) from March 2010 until April 2020.

We use the narrowband generated ToAs published in the North American Nanohertz Observatory for Gravitational Waves (NANOGrav) 15 yr dataset \citep{2023ApJ...951L...9A} where a separate ToA is measured for each frequency channel. The ToA creation follows a standard approach which uses up to 30-minute time-averaged cleaned and calibrated profiles. All profiles for a given receiver band are aligned, weighted by its S/N ratio, and summed to create a final integrated pulse profile. Its average profile is then `denoised' using wavelet decomposition and thresholding to create a final standard template. All narrowband ToAs, were then generated using the Fourier-domain algorithm as implemented in the \texttt{PSRCHIVE} program \texttt{pat}. This method determined each ToA and its uncertainty via a least-squares fit for the pulse phase shift between an observed total-intensity pulse profile and the aforementioned standard template profile.

\subsection{MeerKAT telescope}\label{section:meerkat_telescope}
Observations under the PTA programme were obtained using the MeerKAT telescope between April 16th, 2019 and April 11th 2024. These observations were $\sim$ 480 s each and were regularly spaced, with a mean cadence of two weeks. The RelBin observations were longer and were aimed at obtaining good orbital coverage. In particular, the RelBin data set contains multiple 2~hr observations and one 4~hr (MJD 60302.15) observation taken close to and across superior conjunction to improve the significance of a Shapiro delay measurement.

The MeerKAT observations were recorded using the L-band receiver (856 -- 1712 MHz) with the 1K (1024) channelisation mode, using the Pulsar Timing User Supplied Equipment
(PTUSE) backend \citep{2020PASA...37...28B}. The PTUSE outputs coherently dedispersed folded pulsar archives with 1024 phase bins across the pulse period of 7.98 ms.

Standard array calibration is applied before the observations using the MeerKAT science data processing pipeline (SDP). Details of which can be found in \citet{2021MNRAS.505.4483S}. Before April 9th, 2020, offline polarisation calibration was implemented using the steps outlined in \citet{2021MNRAS.505.4483S}. After April 9, 2020, online polarisation calibration was carried out where the Tied Array Beam data stream input to PTUSE produces polarisation calibrated L-band pulsar data products.

All MeerTime observations were reduced using the \textsc{Meerpipe} pipeline\footnote{\url{https://github.com/OZGrav/meerpipe/}}, which produces cleaned archive files using a modified version of coastguard \citep{2016MNRAS.458..868L} with varying decimation (using standard \texttt{PSRchive/pam} commands). The archives used in this analysis uses the output products with 1024 frequency channels across the inner 775.75 MHz of MeerKAT L-band, 59 sub-integrations and full Stokes information.
To increase the S/N per ToA, we reduced the channelisation to eight frequency channels for all archives and all longer-duration observations were decimated to have a minimum integration length of 3600 s.

Before template creation, all visible residual RFI were removed manually using $\texttt{pazi}$. All observations were then added in time using $\texttt{PSRadd/PSRCHIVE}$. A MeerKAT multi-frequency template was created by decimating the added data to a single subintegration, eight frequency channels, and four Stokes polarisations using the $\texttt{pam/PSRCHIVE}$ command. Analytic frequency-resolved standards were created from this high S/N template by applying wavelet smoothing using $\texttt{PSRmooth/PSRCHIVE}$.
This analytic template was subsequently used to apply matrix template matching \citep{2004ApJS..152..129V, 2013ApJS..204...13V} for ToA creation using $\texttt{pat/PSRCHIVE}$ on the individual MeerKAT data products described in Sect. \ref{section:meerkat_telescope}.
\newline
Lastly, we manually removed individual ToAs with
uncertainties larger than 52 times the RMS ($\sim$3.2 $\mu$s) from all data sets (after visual inspection that confirmed that the large ToA uncertainties were due to low S/N detections).

The time-integrated, frequency-resolved standard profiles from the MeerKAT data are shown in Fig. \ref{Fig:pol}.

\begin{figure}
    \centering
    \includegraphics[width=\columnwidth, trim={50 50 50 50}]{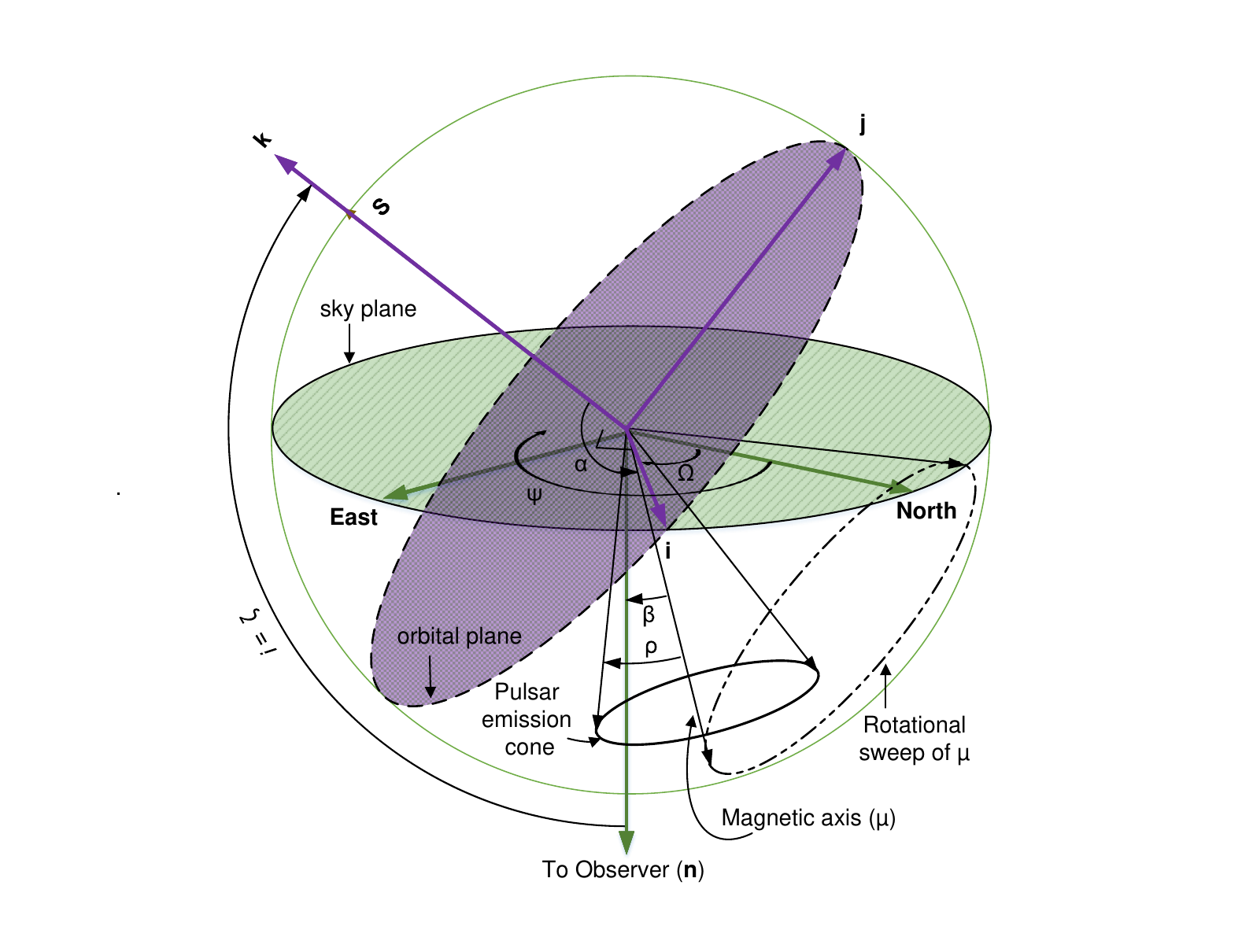}
    \caption{Definition of angles and conventions used in this paper. Throughout this paper, we adopt the “observer’s convention” for all geometric quantities. The fundamental reference plane is in green, and the orbital plane is in purple, with the corresponding unit vectors coloured the same way. \textbf{S} denotes the spin angular momentum, which is aligned with the orbital angular momentum. The magnetic axis ($\mu$) is misaligned from \textbf{S} by the misalignment angle $\alpha$. The pulsar emission cone has an opening angle $\rho$, and is cut through by our line of sight at an impact angle $\beta$. $\zeta = \alpha + \beta$ is the latitude of the spin axis, which is the same as the orbital inclination angle ($i$) for spin-aligned systems.}
    \label{fig:angles}
\end{figure}

\section{Pulsar geometry}\label{Section3}
Throughout this paper, we have used the ``observer's convention'' to quote angles and vectors, unless otherwise stated. These angles are defined in Fig. \ref{fig:angles}, where the position angle and the longitude of the ascending node ($\Omega$) increase counter-clockwise on the plane of the sky (starting from north) and the orbital inclination, $i$, is defined as the angle between the orbital angular momentum and the vector from the pulsar to the Earth.

The emitted electromagnetic waves from a pulsar are polarised along the magnetic field lines, which point radially outwards along the pulsar’s emission cone. As the pulsar's beam moves across the line of sight, the observer sees these magnetic field lines under an ever-changing angle. This effect can be viewed as a variation in the position angle of the linear polarisation (PA; $\Psi$) of the pulse profile across the pulsar longitude. Under ideal assumptions where the magnetic field is assumed to be dipolar and the plane of linear polarisation rotates rigidly with the pulsar, the PA variation results in an S-shaped swing.

We can infer the geometry of J1455$-$3330 from the highly resolved swing of the polarisation angle across the main pulse using the Rotating Vector
Model \citep[RVM;][]{1969ApL.....3..225R}. The RVM describes $\Psi$ as a function of the pulse phase, $\phi$, depending on the magnetic inclination angle, $\alpha$, and the viewing angle, $\zeta'$, which is the angle between the line-of-sight vector and the pulsar spin and can be described as,

\begin{equation}\label{eq:1}
    \Psi =  \Psi_0 + \arctan{\left( \frac{\sin{\alpha} - \sin{(\phi_0 - \phi)}}{\sin{\zeta'} \cos{\alpha} - \cos{\zeta'} \sin{\alpha} \cos{(\phi_0 - \phi})}\right)}.
\end{equation}

Note that all angles are expressed in RVM/DT92 convention; therefore, $\zeta$ and $i$ in the observer's convention are related to $\zeta'$ by $i\sim\zeta=180-\zeta'$.

The RVM does not always describe the polarisation properties of a pulsar (particularly MSPs) well. However, in cases where the RVM does apply, it can be a valuable tool for breaking the sense of the inclination angle obtained from Shapiro delay measurements, enabling a more precise determination of the 3D orbital orientation.

\begin{figure}[h!]
   \centering
   \includegraphics[width=\columnwidth]{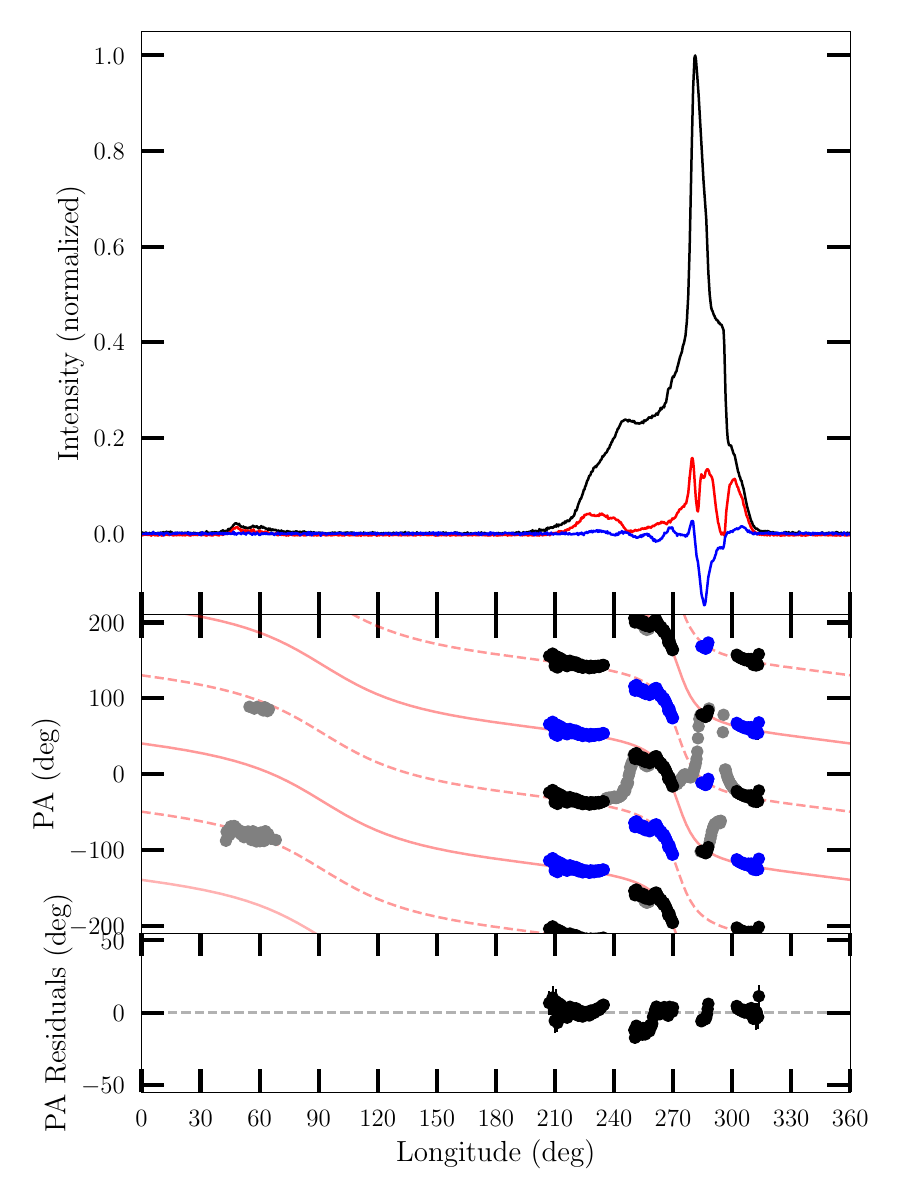}
      \caption{Polarisation profile of J1455$-$3330 obtained from integrating 32.2 hours of observations with the MeerKAT L-band receiver. The black, red, and blue lines in the top panel indicate the total intensity, linear, and circular polarisation fractions, respectively. The middle panel shows the evolution of the position angle (PA) across the pulsar’s phase. The PA
exhibits the characteristic swing as well as some phase jumps. The red
solid line corresponds to the Rotating Vector Model (RVM) fit to the PA, and the dashed line shows the RVM solution separated by 90$^{\circ}$ from the main fit to include the jumped PA values (blue dots).}
         \label{Fig:pol}
\end{figure}

The top panel of Fig. \ref{Fig:pol} shows the polarisation profile of J1455$-$3330 as recorded with the MeerKAT L-band receiver and corrected for the rotation measure given in \citep{2022PASA...39...27S}, the middle panel of Fig. \ref{Fig:pol} shows the the evolution of the position angle (PA) across the pulsar’s phase, and the bottom panel shows the resulting residuals. The PAs are measured in the observer’s convention and exhibit sudden jumps at the edges of the main pulse (MP), coincident with sharp drops in the total linear polarisation. These characteristics can be explained by orthogonal polarisation modes \citep[OPMs;][]{1975PASA....2..334M} and are thought to arise from the propagation effects in the pulsar magnetosphere, or may be intrinsic to the pulsar's emission \citep{1997A&A...327..155G}.

We determine the RVM parameter posteriors in their joint parameter space following the method outlined in \citet{2019MNRAS.490.4565J} using the polarisation angle measurements obtained from all black and blue data points in the middle panel of Fig. \ref{Fig:pol} (from the MeerKAT observations). The model accounts for the possibility of orthogonal polarisation mode jumps and includes the corrected values in the fit. Points that deviate from the RVM-like swing, shown as grey points, are removed. The red solid line corresponds to the RVM fit to the PA, and the dashed line shows the RVM solution separated by 90$^{\circ}$ from the main fit to include the jumped PA values (blue dots).

We performed  six different fits (which are summarised in Table \ref{table:prior}) with varying priors on $\zeta'$: (1) $\zeta'<90^\circ$; (2) $\zeta'>90^\circ$, (3) $53^\circ<\zeta'<70^\circ$ and (4) $110^\circ<\zeta'<127^\circ$. The priors of Runs 3 and 4 reflect the constraints on $\zeta$ imposed from (both senses) of the inclination angle measured from timing (See Section. \ref{sec_5.9}).
A fifth and six fit reflect recent considerations presented by \citet{kj25}, who proposed that the weak pulse component preceding the main pulse by about 150 deg (see Fig. \ref{Fig:pol}) is not originating from the polar cap but from beyond the light cylinder. In this case, its corresponding PA values should be ignored by a fit of the RVM.

\begin{table*}
\caption{$\alpha$ and $\zeta'$ priors and posteriors, log likelihood, and implied inclination angle $i$, for six different RVM model fits.\label{table:prior}}
\centering
\small
\begin{tabular}{lcccccc}
\hline
\hline
\noalign{\smallskip}
 & Run 1 & Run 2 & Run 3 & Run 4 & Run 5 & Run 6 \\
 \noalign{\smallskip}
\hline
\noalign{\smallskip}
Weak component included & Y & Y & Y & Y & N & N \\
$\alpha$ prior ($^\circ$)         & U(0–180) & U(0–180) & U(0–180)   &  U(0-180)    &  U(0-180)&  U(0-180) \\
$\zeta'$ prior ($^\circ$)         & $<90$   & $>90$    & 53 -- 70   & 110 -- 127   & 53 -- 70 & 110-127 \\
Posterior $\alpha$ ($^\circ$)     & 79.0(4)    & 106.1(3)  & 65.4(2)    & 106.1(4)     &  43.9(2) &  105.0(9) \\
Posterior $\zeta'$ ($^\circ$)     & 82.8(4)   & 113.5(4) & 70(5)      &  113.5(4)    &  70(5)   &  112.4(9) \\
Log Likelihood                    & $-626$   & $-330$   & $-785$     & $-330$      &  $-651$   &   $-293$\\
Implied $i$ ($^\circ$)            & 151(5)   & 66.5(2)  & 110(5)     & 66.5(4)    &  110(5)   &   67.6(9)  \\
\noalign{\smallskip}
\hline
\end{tabular}
\end{table*}

\begin{figure}[h!]
   \centering
   \includegraphics[width=\columnwidth]{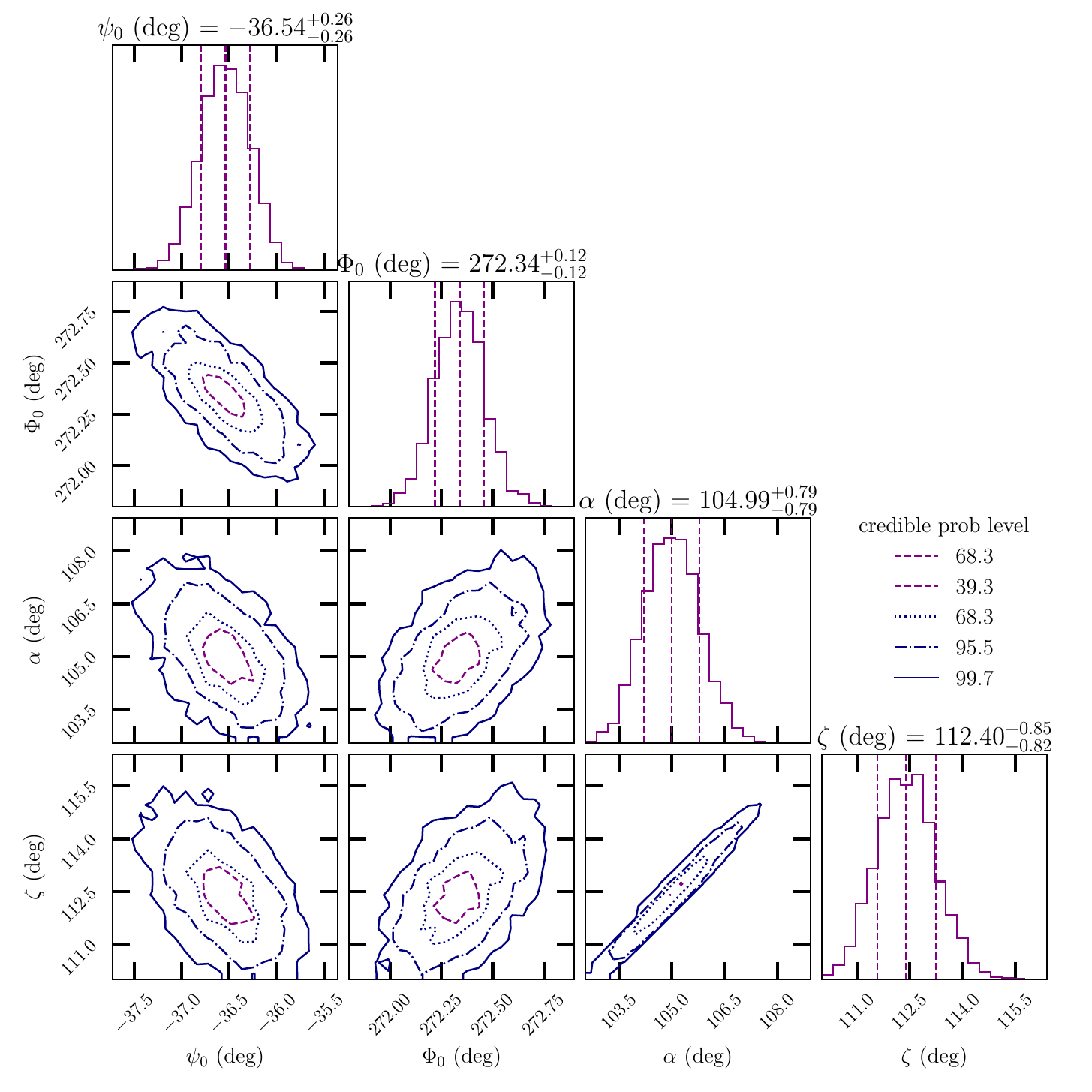}
      \caption{Corner plot showing the posterior distributions from run 2, fitting the Rotating Vector Model (RVM) to the MeerKAT observed position angle variation.}
         \label{Fig:polcorner}
\end{figure}

An unconstrained fit to the entire profile shows that Run 2 is the best-fit model, implying an orbital inclination angle of approximately $67^\circ$.
This can be compared to the timing results discussed later (see Sect. 5.9), as the viewing angle is consistent with the range implied by the sine of the orbital inclination angle ($\sin i$). Constraining the fit of the parameter zeta to a similar range yields the same result for Run 4. In contrast, Run 3's solution attempts to converge with that of Run 1. We also tested solutions following the arguments by Kramer \& Johnston by ignoring the weak component in Runs 5 \& 6. Run 6 yields a very similar result as in Run 2 and 4, as shown in Fig. \ref{Fig:polcorner}.
Keeping in mind the caveats associated with the RVM model in general, see e.g. \citet{2019MNRAS.490.4565J}, the results suggest $\alpha$~=~105.0(9)$^{\circ}$, and $\zeta'$~=~112.4(9)$^{\circ}$, quoting the 68\% confidence levels on the posteriors. This implies an outer line-of-sight of the beam with a pole crossing as derived by the RVM of  $\phi_0 = 272^{\circ}$, placing the fiducial plane right under the MP's central peak as usually expected.

\section{Timing analyses}\label{Section4}
In total, we have 16,461 ToAs in the dataset, 364 ToAs from the Lovell telescope data, 4,288 ToAs from the NRT, 10818 ToAs from the GBT, and 991 ToAs from MeerKAT.

Timing analyses were made using the \texttt{tempo2}\footnote{\url{https://bitbucket.org/psrsoft/tempo2/src/master/}} \citep{2006MNRAS.369..655H, 2006MNRAS.372.1549E} software package.

The different data sets were combined using an arbitrary phase offset or ``JUMP'' between each of them, using the MeerKAT data as the reference data set. These jumps account for additional time delays due to different backend instruments and geographical locations and are an arbitrary phase offset between each telescope dataset.

All ToAs acquired and generated were first transformed into
TT(BIPM2021)\footnote{\url{ https://webtai.bipm.org/ftp/pub/tai/ttbipm/TTBIPM.2021}}, which is a version of terrestrial time as defined by the International Astronomical Union (IAU). Thereafter, these ToA's are converted into ToAs at the Solar System barycentre using the DE438 Solar System ephemeris of the Jet Propulsion Laboratory \citep[JPL]{2021AJ....161..105P}.

The initial orbital, astrometric and pulsar parameter estimates were found using the ELL1H orbital model implemented in the \texttt{tempo2} software \citep{2010MNRAS.409..199F}. This model is based on the ELL1 model \citep{2001MNRAS.326..274L}, which is designed to avoid the extreme correlation between the epoch of periastron ($T_0$) and the longitude of periastron at $T_0$ ($\omega$) for low-eccentricity orbits, with the $xe^2$ terms \citep{2019MNRAS.482.3249Z} added for accuracy; this is especially important for wide systems like J1455$-$3330. 

The ELL1H model mitigates the correlation between the 2 PK parameters that quantify the Shapiro delay measurement (range, $r$ and shape, $s$) by re-parameterizing the Shapiro delay with 2 different PK parameters, the orthometric amplitude ($h_3$) and orthometric ratio ($\varsigma$) \citep{2010MNRAS.409..199F}.

This model, however, is only able to account for the secular variation of the projected semi-major axis ($\dot{x}$) and not the full set of kinematic contributions to the variation (see \citealt{1995ApJ...439L...5K,1996ApJ...467L..93K}). The model cannot discriminate between the multiple solutions for the orbital orientation of the system given by the $\dot{x}$ and inclination angle. Therefore, we also use the T2 orbital model \citep{2006MNRAS.372.1549E}, which is based on the DD model \citep{1986AIHPA..44..263D} and self-consistently accounts for all kinematic contributions to the orbital and post-Keplerian parameters. All kinematic contributions to the changes of orbital orientation with respect to our line of sight are calculated internally from the orbital orientation of the system given by the longitude of the ascending node ($\Omega$) and orbital inclination ($i$). In particular, the T2 model does not separately fit for the variation of the projected semi-major axis of the pulsar's orbit ($\dot{x}$).

\begin{figure*}
   \centering
   \includegraphics[width=0.85\textwidth]{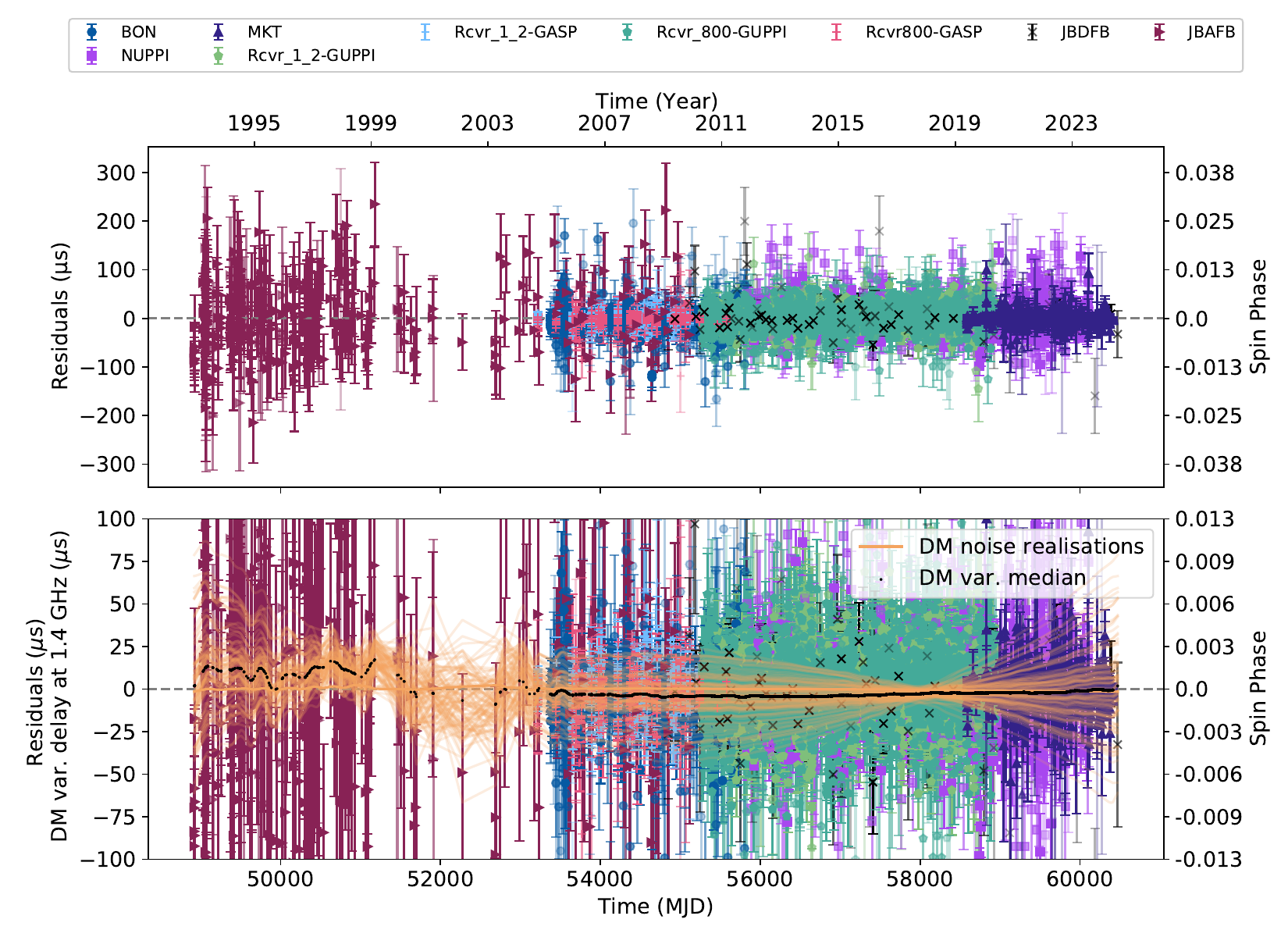}
   \includegraphics[width = 0.9\textwidth]{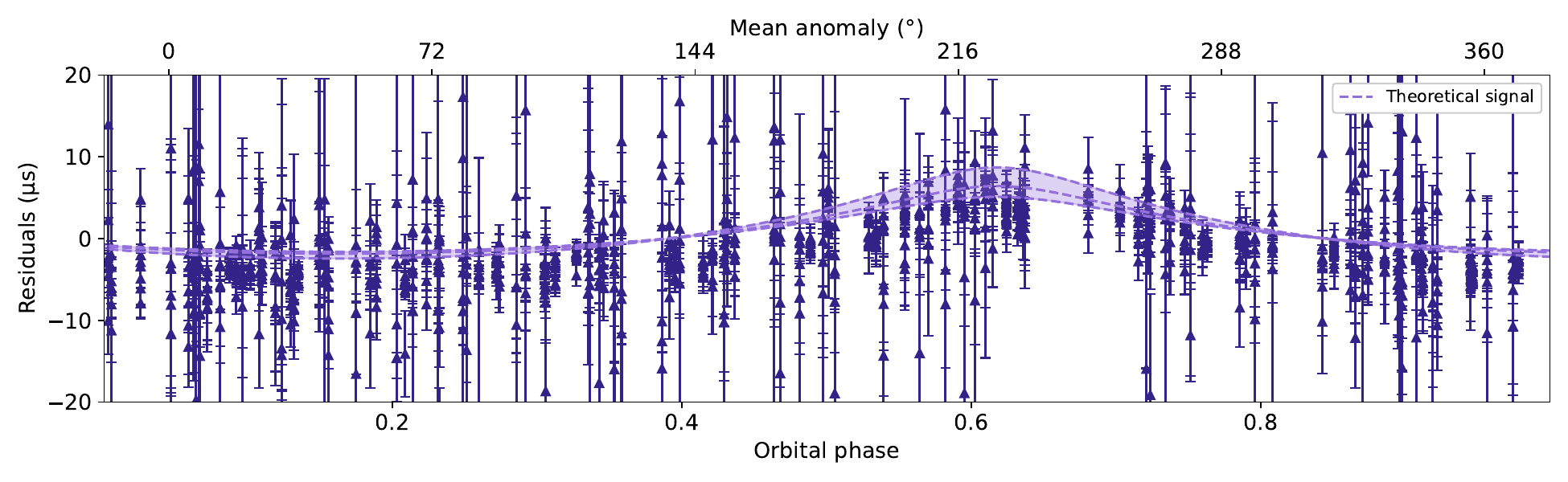}
      \caption{Timing residuals across epochs (top, middle) and orbital phase (bottom). Top: We obtain a weighted rms of 3.201 $\mu$s after applying the best-fit values using the T2 timing and noise model described in Sect. \ref{Section4}.  Middle: Post-fit timing residuals without subtracting the DM noise model, with the time domain realisations of the 100 parameter DM noise model overlaid as orange lines, and black dots showing the median across all the DM model realisations at each ToA.
      Bottom: MeerKAT residuals as a function of orbital phase, where the orbital phase is measured from the longitude of periastron ($\omega=223.47^\circ$). Superior conjunction = $T_{\rm asc}$ + $90^\circ$, occurs at orbital phase = 0.63. A Shapiro delay signal is discernible at orbital phase 0.63 when setting $M_{\rm c} = 0$ while keeping all other parameters fixed. We overplot the expected theoretical signal based on the best-fit inclination and companion mass values of the full dataset in purple. The line width indicates the 1$\sigma$ deviations in companion mass and inclination angle.}
         \label{Fig:Residuals}
   \end{figure*}

To understand and eliminate stochastic noise variations in our dataset, we used the $\texttt{temponest}$ \citep{2014MNRAS.437.3004L} plugin to \texttt{tempo2}, which is a Bayesian parameter estimation tool used to perform non-linear fits of the T2 timing model to the data. We chose the best-fit noise model to describe the data by testing four different noise models. This selection process includes a white-noise only model, a white-noise plus red-noise model, a white-noise plus DM-noise model, and a white-noise plus DM-noise plus red-noise model.
In each of these scenarios, a white-noise model is described by EFAC and EQUAD (which scales the uncertainties of the ToAs linearly and quadratically, respectively), a DM-noise model is described by a chromatic power-law model, and a stochastic achromatic power law model describes a red-noise model. For each model, we provide uniform priors centred on the initial best-fit \texttt{tempo2} parameter $\pm$ 40 $\sigma$, where $\sigma$ is the associated \texttt{tempo2} uncertainty. For a select set of parameters, we
provided physically motivated priors in two separate \texttt{temponest} runs. This allowed us to explore multiple plausible geometries of the system while avoiding premature convergence to a single local solution.
The first run's $\Omega$ was set to cover the range of possible values 0-360$^\circ$ and $i$ was set to cover the range of values 0-90$^{\circ}$. The second run's $\Omega$ was set to cover 270-450$^\circ$ (-90-90$^\circ$), and $i$ was set to cover the range of possible values 90-180$^\circ$. The parallax, $\varpi$, was set to range from 0.3 to 3, and $M_{\rm c}$ from 0.1 to 1.4 $\rm M_{\odot}$ for both \texttt{temponest} runs.

We performed comparisons between the
models using the Nested Importance Sampling Global Log-Evidence and we find the strongest evidence for a white-noise plus DM-noise model. All DM effects are well-modelled by including the DM1 and DM2 timing parameters. These parameters describe the coefficients to the first- and second-order DM derivatives and are expressed as a DM Taylor series expansion.

For the remainder of the paper, we use the outcomes of
the \texttt{temponest} posterior distributions, which include DM-noise parameters only. The amplitude and power-law spectral index of the DM-noise are provided in Table \ref{Tab:3}. 

A recent investigation of the influence of different timing models on measured parameters of J1455$-$3330 using the 15 yr NANOGrav dataset \citep{2025arXiv250603597L} found a preference for a linear trend in DM plus a fixed solar wind density over a variety of models (DMX model, a linear trend in DM plus a fixed solar wind density, a quadratic trend in DM plus a fixed solar wind density, and a linear trend in DM plus a varying solar wind density). Although the parallax varies across models, the value obtained using the most favoured model agrees with the value obtained from our analyses of the full dataset. Furthermore, we use the time domain realisation methods of the La Forge GitHub repository\footnote{https://github.com/nanograv/la_forge}
\citep{2020zndo...4152550S} and overplot the resultant realisations on the post-fit timing residuals (without including the DM noise model) as orange lines on the middle panel of Fig. \ref{Fig:Residuals}. The noise model aligns with the visible trends in the data, further demonstrating the validity of the noise model.

\section{Results}\label{Section5}
\begin{figure*}
   \centering
   \includegraphics[width=\textwidth]{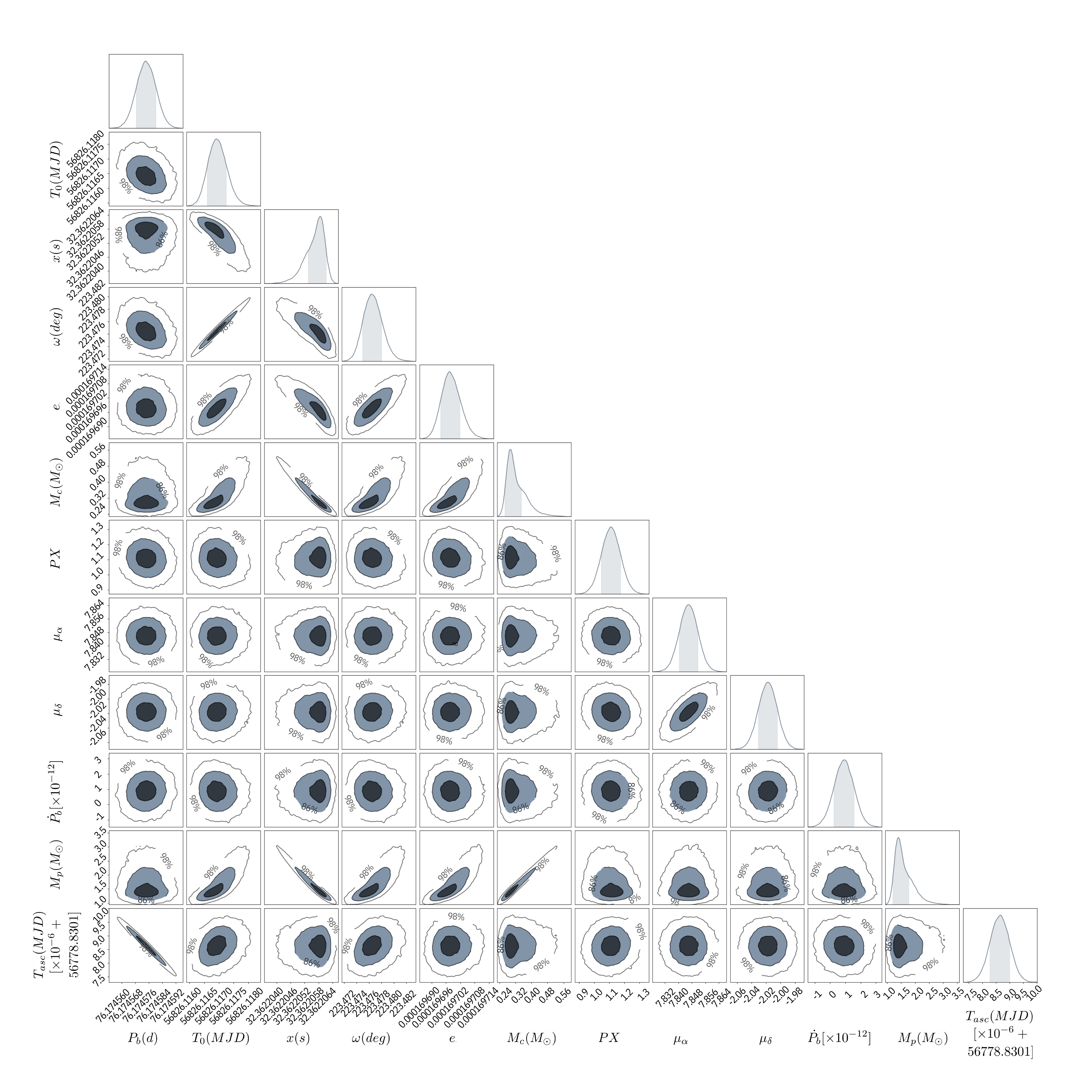}
      \caption{Output posterior distribution for the relevant orbital and post-Keplerian parameter subset of timing parameters for J1455$-$3330. These were obtained from \texttt{temponest} using the T2 orbital model which includes a DM only noise model. The plot was generated using the \texttt{chainconsumer} package. Full details of the parameters are provided in Table \ref{Tab:3}. The obtained pulsar mass ($M_\rm p$) distribution was computed using the mass function and the posterior distributions on $M\rm c$,$i$, $x$, and $P_\rm b$. The 1D marginalised posterior distributions for each parameter are shown on the diagonal subplots and the shaded region indicates the 1-$\sigma$ credibility interval. The 2D
contours on the off-diagonal subplots show the correlation between pairs of parameters, where the contours mark the 39\%, 86\% and 98\% credibility regions respectively.}
         \label{Fig:temponest}
   \end{figure*}

In this section, we present the complete set of spin, astrometric, binary and derived parameters using the T2 model. These are shown in Table \ref{Tab:2}. 
The final \texttt{temponest} solution is a good description of the data and the timing residuals obtained after comparing individual ToAs with the model are shown in Fig. \ref{Fig:Residuals}. The lack of any visible trends in the residuals and the low weighted rms of $\sim$ 3.19~$\mu$s further show how well the model fits the data. 
To visualise the \texttt{temponest} T2 posterior distributions with corner plots, we used the $\texttt{chainconsumer}$ library \citep{2016JOSS....1...45H} to plot the 1D and 2D posterior distributions of the measured parameters. Fig. \ref{Fig:temponest} shows the resulting output for a subset of timing parameters of interest. In this plot, we used the DM-noise model in \texttt{temponest} with 5000 live points to produce well-sampled distributions. The dark blue, medium blue and white contours on the 2D off-diagonal plots show the 39\%, 86\%, and 98\% credibility regions respectively. The shaded region on the 1D diagonal plots represents the 68.27$\%$ credibility region. 

We briefly highlight each of the measured and derived parameters in the following subsections. These parameters consist of proper motion, parallax, spin parameters, orbital period derivative ($\dot{P}_{\rm b}$), rate of advance of periastron ($\dot{\omega}$), Shapiro delay, change in projected semi-major axis ($\dot{x}$), annual orbital parallax, and finally the constraints on the component masses and 3D orbital geometry.

\subsection{Position and proper motion}
We provide an updated position and proper motion for J1455$-$3330. The updated J2000 position obtained from timing is right ascension,~$\alpha$~=~ 14:55:47.973129(2), and declination,~$\delta$~=~ $-$33:30:46.39040(6). From the measured proper motion in right ascension \mbox{($\mu_{\alpha} = 7.846(6)\, \rm  mas\, yr^{-1})$} and proper motion in declination ($\mu_{\delta} = -2.02(1)\, \rm mas\, yr^{-1}$), we obtain a total proper motion $\mu_{\rm T} = 8.101(6)\, \rm mas \, yr^{-1}$. The corresponding position angle of the proper motion, $\theta_{\mu}$ = tan$^{-1}$($\mu_{\alpha}/\mu_\delta$) is 284.42(9)$^{\circ}$. 

\begin{table*}[t]
	\centering
	\caption{J1455$-$3330 timing parameters obtained from \texttt{temponest}.}
	\label{Tab:2}
         \begin{tabular}{ll}
            \hline
            \multicolumn{2}{c}{\rm Observations and data reduction parameters}\\
		\hline
            Timing model\dotfill& T2\\
            Solar System ephemeris\dotfill&DE438\\
            Timescale\dotfill&TT(BIPM2019)\\
            Reference epoch for period, position and DM (MJD)\dotfill&56893\\
            Solar wind and electron number density, n (cm$^{-3}$)\dotfill&4\\
            Weighted rms residual ($\mu$s)\dotfill & 3.19\\
            \hline
            \multicolumn{2}{c}{Spin and astrometric parameters}\\
            \hline
            Right ascension, $\alpha$ (J2000, h:m:s) \dotfill& 14:55:47.973129(2)\\
             Declination, $\delta$ (J2000, d:m:s)\dotfill& $-$33:30:46.39040(6)\\
             Proper motion in $\alpha$, $\mu$$\alpha$ (mas yr$^{-1}$)\dotfill&  $ 7.846(6)$ \\        
             Proper motion in $\delta$, $\mu$ $\delta$ (mas yr$^{-1}$)\dotfill& $-2.02(1)$ \\
             Parallax, $\varpi$ (mas) \dotfill&    $1.11(6)$\\
             Spin frequency, $\nu$ (Hz) .\dotfill& 125.20024512394019(7)\\
             Spin-down rate, $\dot{\nu}$ (10$^{-16}$ Hz s$^{-1}$)\dotfill& $-3.809561(6)$ \\
             Dispersion measure, DM (cm$^{-3}$ pc)\dotfill& 13.5699(1)\\
             First Derivative of DM, DM1 ($10^{-5}\, \rm cm^{-3}\, pc \, yr^{-1}$) \dotfill& 5(2)\\
             Second Derivative of DM, DM2 ($10^{-6}\, \rm cm ^{-3}\, pc\, yr^{-1}$)\dotfill& 8(3) \\ 
             \hline
             \multicolumn{2}{c}{Derived parameters}\\
             \hline
             Galactic longitude, $l$ ($^{\circ}$) \dotfill& 330.72177524\\
             Galactic latitude, $b$ ($^{\circ}$) \dotfill &  22.56224922\\
             Total proper motion, $\mu_{\rm T}$ (mas yr$^{-1}$)\dotfill & 8.101(6)\\
             DM-derived distance (NE2001), $d$ (kpc)\dotfill & 0.52737\\
             DM-derived distance (YMW16), $d$ (kpc)\dotfill & 0.68418\\
             Parallax derived distance, $d$ (kpc)\dotfill & 0.90(5)\\
             Parallax derived distance including EDSD prior, $d_{\rm PSR}$ (kpc)\dotfill & 0.90(5)\\
             Spin period, $P$ (ms)\dotfill &              7.987204809463945(5)\\
             Spin period derivative, $\dot{P}$ (10$^{-20} \, \rm s \,s^{-1}$)\dotfill &2.430325(4)\\
             Total kinematic contribution to $\dot{P}$, $\dot{P}_{\rm k}$ ($10^{-21}\, \rm  s\, s^{-1}$)\dotfill &1.8(1)\\      
             Intrinsic spin period derivative $\dot{P}_{\rm int} (10^{-20} \, \rm s\, s^{-1})$\dotfill & 2.25(1)\\
             Inferred surface magnetic field, $B_{s} (10^{8}\, \rm G)$ \dotfill& $\sim 4.23$\\
             Inferred characteristic age, $\tau_{c} \, \rm (Gyr)$ \dotfill& $\sim$ 5.63\\
             Inferred spin-down luminosity, $\dot{E} (10^{33} \, \rm erg s^{-1})$\dotfill&$\sim$ 1.736(8)\\
            \hline
            
	\end{tabular}
  \tablefoot{Numbers in parentheses are the nominal 1-$\sigma$ maximum symmetric statistical uncertainties on the last quoted digits. We used the ELL1H model to fit for the Shapiro delay and $\dot{x}$. In the second half of the table, we present quantities derived from the fit values. The last three values are calculated from $\dot{P}_{\rm int}$ after correcting for the kinematic effects.}
\end{table*}

\subsection{Parallax}

Combining the estimated DM with models of the electron distribution of the Galaxy, we infer a distance to the pulsar between $\sim$0.527 kpc \citep[from the NE2001 model]{2002astro.ph..7156C} and $\sim$ 0.684 kpc \citep[from YMW16 model]{2017ApJ...835...29Y}. 

In this case, we have a direct measurement of the pulsar parallax, $\varpi~=~1.11(6)\, \rm mas$, which can be inverted to provide a distance estimate of 0.90(5) kpc. However, this relation is prone to the Lutz-Kelker bias \citep{1973PASP...85..573L} where parallax measurements are overestimated because they do not properly account for the larger volume of space that is sampled at smaller parallax values. We correct for this bias using a scaled probability density
function. Following \cite{2012ApJ...755...39V} and \cite{2021MNRAS.501.1116A} we infer a probability density function for the distance to the system using:
\begin{equation}
P(d_{\rm PSR}|\varpi) = \frac{1}{2L^{3}} e^{-d_{\rm PSR}/L} d^{2} 
\exp\left( -\frac{\left( \frac{1}{d_{\rm PSR}} - \varpi \right)^2}{2\sigma_{\varpi}^2} \right)
\end{equation}
We assume that the parallax measurement, $\varpi$ is normally distributed about the true parallax, $d^{-1}$, with a dispersion $\sigma_{\varpi}$. The $e^{-d_{\rm PSR}/L}d^{2}L^{-3}$ term accounts for a Lutz-Kelker bias \citep[L-K bias]{1973PASP...85..573L} with an exponentially decreasing stellar density almost constant for $d$ $\ll$ $L$). L can be considered as a characteristic length scale, which is set to 1.35 kpc following \cite{2021MNRAS.501.1116A}. 
The estimate of the pulsar distance ($d_{\rm PSR}$) corrected for the L-K bias from the measured parallax is $0.90^{+0.05}_{-0.06}$~kpc.
Combining $d_{\rm PSR}$ with $\mu_{\rm T}$, we obtain a heliocentric transverse velocity $V_{\rm T} = 34.5 \pm 1.9 \,$  km $\rm s^{-1}$.

\subsection{Spin parameters}\label{spin}
The observed spin period of the pulsar includes both the intrinsic spin period and the kinematic contributions.
\begin{equation}
    \begin{aligned}
         \left( \frac{\dot{P}}{P} \right)^{\rm obs} &= \left( \frac{\dot{P}}{P} \right)^{\rm int} + \left( \frac{\dot{P}}{P} \right)^{\rm Shk} + \left( \frac{\dot{P}}{P} \right)^{\rm Gal}, \\
    \end{aligned}
\end{equation}
where $\dot{P}_{\rm int}$ is the intrinsic spin-down rate, P is the pulsar's spin period, $\dot{P}_{\rm obs}$ is its observed spin-down rate. The kinematic effects are caused by the combined effect of; (1) $\dot{P}_{\rm Shk}$ which is the Shklovskii effect  \citep{1970SvA....13..562S}, and (2) the $\dot{P}_{\rm Gal}$ which is the acceleration of the binary in the gravitational field of the Milky Way due to differential rotation.
$(\dot{P}/{P})^{\rm Shk}$ depends on $\mu_{\rm T}$ and $d_{\rm PSR}$ as follows:
\begin{equation}
          \left(\frac{\dot{P}}{P}\right)^{\rm Shk} = 2.43 \times 10^{-21} \left( \frac{\mu_T}{\text{mas yr$^{-1}$}} \right)^{2} \left( \frac{d_{\rm PSR}}{\text{kpc}}\right)\\
          \label{eqn2}
\end{equation}
$(\dot{P}/P)^{\rm Gal}$ represent the Galactic acceleration which is the difference between the accelerations of the pulsar binary system and the Solar System in the gravitational field of the Galaxy, projected along the line of sight from the Solar System to the pulsar binary system. This Galactic acceleration includes contributions from both the Galactic rotation and vertical accelerations relative to the disk of the Galaxy. Following \citet{1991ApJ...366..501D}, \citet{1995ApJ...441..429N}, 
\citet{2009MNRAS.400..805L}, and \cite{2018ApJ...868..123P}:
\begin{subequations} \label{eqn3}
\begin{align}
\left(\frac{\dot{P}}{P}\right)^{\rm Gal} &= \left( \frac{\dot{P}}{P} \right)^{\rm Gal, Pl} + \left( \frac{\dot{P}}{P} \right)^{\rm Gal, Az} \text{where, }\\
\left( \frac{\dot{P}}{P} \right)^{\rm Gal, Pl} &=
    -\frac{\Omega_{\odot}^2}{c R_{\odot}} \left( \cos{l} +\frac{\beta}{\beta^2 + \sin^2 l
} \right) \cos{b}\\
\left( \frac{\dot{P}}{P} \right)^{\rm Gal, Az} &= \frac{-K_z |\sin{b}|}{c}
\end{align}
\end{subequations}

where $\beta$ = $(d_{\rm{PSR}}/R_{\odot}) \cos b - \cos l$ and $z_{\rm{kpc}} = d_{\rm PSR} |\sin b|$ in kpc. $R_{\odot} = 8.28(3) \, \rm kpc$ is the distance to the galactic centre, and $\Omega_\odot = 241(4)\, \rm km \, s^{-1}$ is the galactic rotation velocity \citep{2021A&A...647A..59G,2021A&A...654A..16G}.
K$_{\rm z}$/c is the vertical component of Galactic acceleration, and is given by \citet{2017MNRAS.465...76M}:
\begin{equation}
    \frac{K_z}{c} \left[s^{-1}\right] = -1.08\times10^{-19}\left( 0.58 + \frac{1.25}{(z_{\rm{kpc}}^{2}+0.0324)^{1/2}} \right) z_{\rm{kpc}}
    \label{eqn4}
\end{equation}

Using the distance estimated from parallax, the Shklovskii effect $(\dot{P}/P)^{\rm Shk} = 1.15(7)\times10^{-21}$ s$^{-1}$ and the Galactic acceleration $(\dot{P}/P)^{\rm Gal} = 6.8(3)\times10^{-22}$ s$^{-1}$. Subtracting these two terms from $\dot{P}^{\rm obs}$=$2.430\times10^{-20}$ s s$^{-1}$ we obtain the ‘intrinsic’ variation in the spin period, $\dot{P}^{\rm int} = 2.25(1)\times10^{-20}$ s s$^{-1}$.
This intrinsic spin-down value is very close to the observed spin-down since the Shklovskii and Galactic terms are $\sim$ 
10 - 100 times smaller. Using this $\dot{P}_{\rm int}$, we estimate the surface magnetic field ($B_{s}~=~4.230(9)\times10^{8}~G$), the characteristic age ($\tau_c~=~5.63(3)$~Gyr), and the spin-down luminosity of the system \citep[$\dot{E} = 1736(8)\times10^{30} \, \rm erg \, s^{-1}$,][]{2004hpa..book.....L}.

\begin{table*}[t]
	\centering
	\caption{Binary timing parameters and associated mass and inclination values for J1455$-$3330.}
	\label{Tab:3}
         \begin{tabular}{llll}
		\hline
        \hline
             Orbital model&\multicolumn{2}{c}{T2}&ELL1H\\
             \hline
            \multicolumn{4}{c}{Keplerian parameters}\\
		\hline
            Orbital period, $P_{\rm b}$ (days)\dotfill& $76.174574(6)$&\\
            Projected semi-major axis of the pulsar orbit, $x_{\rm p}$ (s)\dotfill &$32.3622059(4)$&\\
            Epoch of periastron, ($T_0$) (MJD)\dotfill & $56826.1169(3)$ &\\
            Orbital eccentricity, $e$ \dotfill & 0.000169697(4)&\\
            Longitude of periastron at $T_0$, $\omega$ ($^{\circ}$)\dotfill & 223.476(2)&\\
            \hline
             \multicolumn{4}{c}{Post-Keplerian parameters}\\
            \hline
            &TN Run 1&TN Run 2\\
            Orbital Inclination, $i$ prior\dotfill&U(0-90)&U(90-180)\\
            Longitude of the ascending node, $\Omega$ prior \dotfill&U(0-360)&U(270-450)\\
            Orbital period derivative, $\dot{P}_{\rm b}$ (10$^{-13}\, \rm s\,s^{-1}$)\dotfill & $9(7)$&11(7)&\\
            Rate of advance of periastron, $\dot{\omega}$ (10$^{-4}$ $ ^{\circ}$ yr$^{-1}$)
            \dotfill&1.5(1.3)&1.5(1.5)&\\
            Rate of change of orbital semi-major axis, $\dot{x}$ (10$^{-16}\, \rm s \, s^{-1}$) \dotfill&&& $-202(3)$\\
            Shapiro delay amplitude, $h_3$ ($10^{-7}\, \rm s$) \dotfill&&& $3.07^{+0.22}_{-0.26}$\\
            Orthometric ratio, $\varsigma$ \dotfill &&& $0.551^{+0.057}_{-0.054}$\\
            \hline
           \multicolumn{4}{c}{Mass and orbital geometry measurements}\\
            \hline
            Longitude of the ascending node, $\Omega$ ($^{\circ}$) \dotfill &212(12)&$334(12)$\\
            Companion mass, $M_{\rm c}$, (M$_\odot$)\dotfill &$0.293^{+0.056}_{-0.026}$&0.309$^{+0.163}_{-0.026}$\\
            Orbital inclination, $i$ (deg)\dotfill&63(2) & $123(4)$\\
            \hline
            \multicolumn{4}{c}{Noise parameters}\\
            \hline
            DM-noise amplitude (log$_{10}$ $A_{\rm DM}$)\dotfill & -11.73(6)
&-11.71(6)& -11.52(4)\\
            DM-noise spectral index ($\alpha_{\rm DM}$)\dotfill&1.6(2)&1.6(2) &1.3(1)\\
            \hline
             \multicolumn{4}{c}{Derived parameters}\\
            \hline
            Mass function, $f$ (M$_\odot$)\dotfill & 0.00627158492 (21) &\\
            Pulsar mass, $M_{\rm p}$ (M$_\odot$)\dotfill &1.39$^{+0.38}_{-0.18}$&$1.53^{+1.10}_{-0.22}$\\
            Total kinematic contribution to $\dot{P}_{\rm b}$ from Shklovskii effect, $\dot{P}_{\rm b, Shk}$, (10$^{-13} \, \rm s \, s^{-1}$)\dotfill&$9.5(5)$ &\\
            Total kinematic contribution to $\dot{P}_{\rm b}$ from Galactic rotation, $\dot{P}_{\rm b, Gal}$, (10$^{-13}\, \rm  s \, s^{-1}$)\dotfill&$5.61(25)$ &\\
           
            Intrinsic orbital period derivative ($10^{-13}\, \rm s \, s^{-1}$), $\dot{P}_{\rm b, int}$\dotfill& $-6(7)$ & \\
            \hline
            \hline
	\end{tabular}
        \tablefoot{TN Run 1 and TN Run 2 correspond to two separate \texttt{temponest} runs, including different priors for $i$ and $\Omega$. U(x-y) denotes uniform distributions ranging from x to y for the relevant parameters.}
\end{table*}

\subsection{Binary period derivative: $\dot{P}_{\rm b}$}

Using the current timing baseline, we did not obtain a significant detection of the orbital period derivative ($\dot{P}_{\rm b} =  9(7) \times 10^{-13}$ s s$^{-1}$). The expected contributions to $\dot{P}_{\rm b}$ are:
\begin{equation}
    \left( \frac{\dot{P}_{\rm b}}{P_{\rm b}} \right)^{\rm obs} =  \left( \frac{\dot{P}_{\rm b}}{P_{\rm b}} \right)^{\rm GR} + \left( \frac{\dot{P}_{\rm b}}{P_{\rm b}} \right)^{\rm Shk} + \left( \frac{\dot{P}_{\rm b}}{P_{\rm b}} \right)^{\rm Gal}  + \left( \frac{\dot{P}_{\rm b}}{P_{\rm b}} \right)^{\dot{m}} + \left( \frac{\dot{P}_{\rm b}}{P_{\rm b}} \right)^{T},
\end{equation}
where $(\dot{P}_{\rm b}/P_{\rm b})^{\rm obs}$ is the observed orbital period derivative, $(\dot{P}_{\rm b}/P_{\rm b})^{\rm GR}$ is the contribution due to gravitational wave decay, $(\dot{P}_{\rm b}/P_{\rm b})^{\rm Shk}$ and $(\dot{P}_{\rm b}/P_{\rm b})^{\rm Gal}$ are the same kinematic contributions described in Sect. \ref{spin}, $(\dot{P}_{\rm b}/P_{\rm b})^{\dot{m}}$ is the mass loss in the system, and $(\dot{P}_{\rm b}/P_{\rm b})^{\dot{T}}$ is the tidal dissipation of the orbit. We find that the only non-negligible contribution arises from the kinematic terms.

\begin{figure}[h]
   \centering
   \includegraphics[width=\columnwidth]{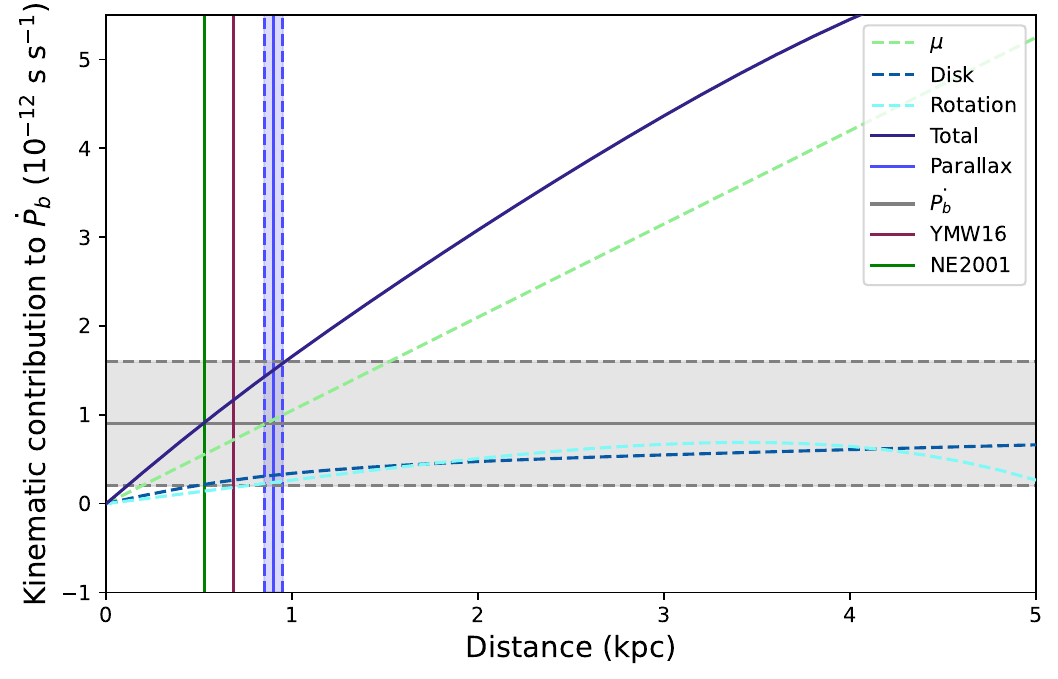}
      \caption{Kinematic contributions to $\dot{P_{\rm b}}$ as a function of distance. The distance constraints from $\varpi$, the NE2001 model and the YMW16 model are shown as vertical lines. The constraint from $\dot{P}_{\rm b, obs}$ is shown as the gray shaded region. The curves show the contributions from the vertical and horizontal acceleration from the Galactic disc, the proper motion of the system, and the total acceleration.}
         \label{Fig:distance_plot}
\end{figure}

From equations \ref{eqn2}, \ref{eqn3}, and \ref{eqn4}, we obtain the following results: $\dot{P}_{\rm b, Shk}$ = $9.5(5)\times10^{-13}$ s s$^{-1}$ and $\dot{P}_{\rm b, Gal}$ = $5.6(3)\times10^{-13}$ s s$^{-1}$. In Fig.~\ref{Fig:distance_plot}, we can see how these quantities and their sum change as a function of the distance.

The observed $\dot{P}_{\rm b}$ is given by $\dot{P}_{\rm b, obs} = 9(7)\times10^{-13}$ s s$^{-1}$, this is shown in Fig.~\ref{Fig:distance_plot} as the gray bar. Within the $\pm 1$-$\sigma$ uncertainties, this matches well the sum of kinematic terms expected for the distance as measured via parallax. As we can see in the figure, the steepness of the curve of the total kinematic effects implies that a more precise value of $\dot{P}_{\rm obs}$ will also provide an independent estimate of the distance, which can be compared with the parallax distance. 

The intrinsic $\dot{P}_{\rm b}$ is given by $\dot{P}_{\rm b, int} = \dot{P}_{\rm b, obs} - \dot{P}_{\rm b, Skk} - \dot{P}_{\rm b, Gal} = -6(7)\times10^{-13}  \, \rm s \, s^{-1}$.
This is expected to originate from the orbital decay of the system caused by gravitational wave emission. Using the orbital parameters, masses (derived from the T2 model) and the relation from \citet{1964PhRv..136.1224P}, we estimate the orbital decay caused by the emission of quadrupolar gravitational waves in GR, $\dot{P}_{\rm b}^{\rm GW} = -5.5(1.2)\times10^{-18}\, \rm s \, s^{-1}$. This is 5 orders of magnitude below the current timing precision, i.e. within measurement precision it is expected to be consistent with zero, as observed.

\begin{figure*}[t]
   \centering
   \includegraphics[width=1.0\textwidth]{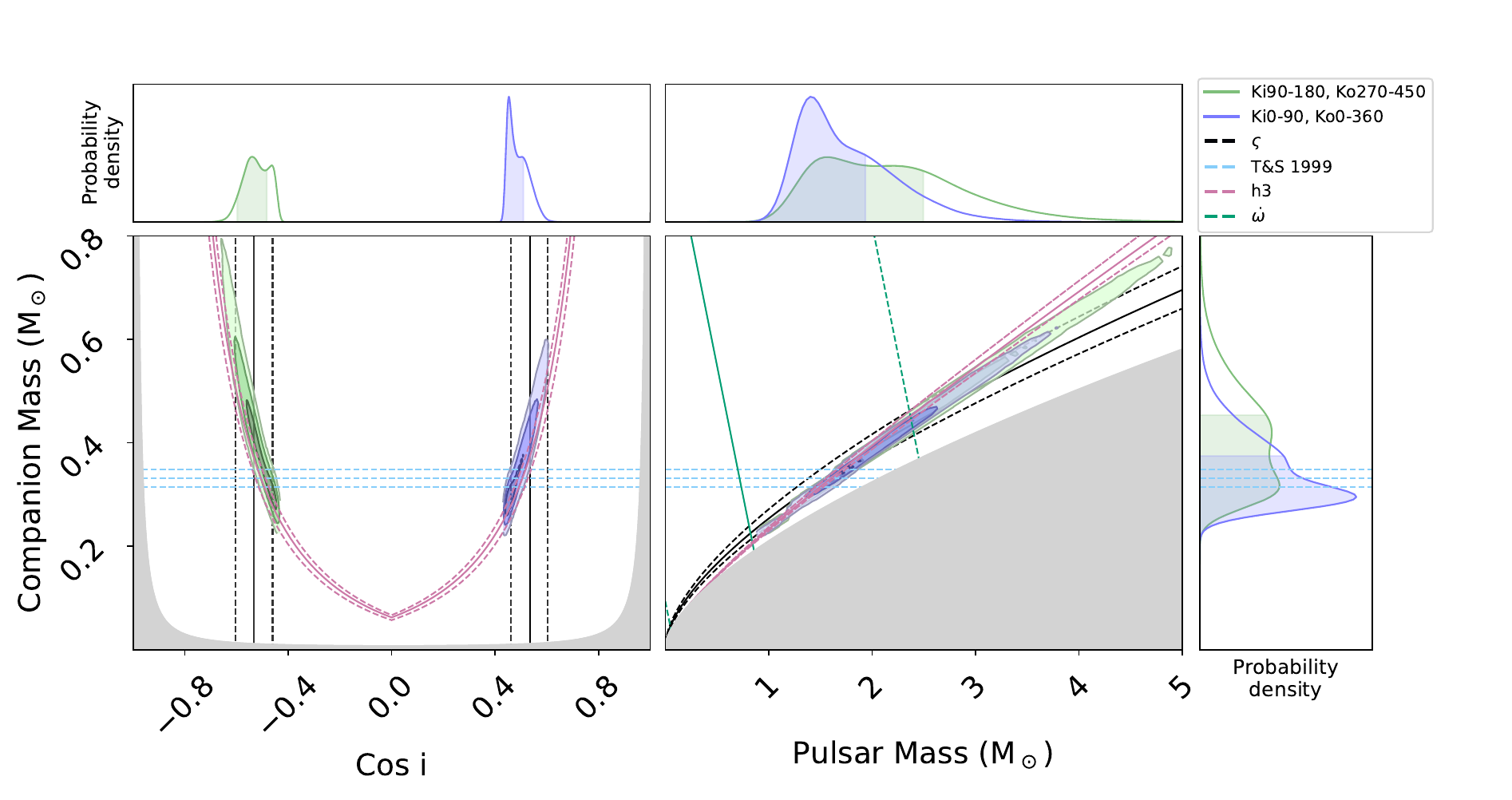}
      \caption{Constraints on the companion mass as a function of the cosine of the orbital inclination (left), and as a function of the pulsar mass (right). On the left plot, the gray region is excluded by the requirement that the pulsar mass must be greater than zero. On the right plot, the mass values in the gray region is excluded by the mass function. The black solid and dashed lines and the pink solid and dashed lines represent the median, and 1 sigma $\varsigma$ and $h_3$ measurements using the ELL1H model respectively. The solid contours enclosing progressively darker shades of purple include 98\%, 86\% and 39\% confidence limits of the 2D probability density function from the T2 timing model solutions calculated by \texttt{temponest} with priors of $i$ ranging from 90 to 180 degrees and $\Omega$ ranging from 270 to 450 degrees. The solid contours enclosing progressively darker shades of purple include 98\%, 86\% and 39\% confidence limit of the 2D probability density function from the T2 timing model solutions calculated by \texttt{temponest} with priors of $i$ ranging from 0 to 90 degrees and $\Omega$ ranging from 0 to 360 degrees. The marginalised posterior probability distributions for $\cos{i}$, $M_\rm c$ and $M_\rm p$ are displayed for each axis. The shaded region on the marginalised plots at the top and the right represents the 68\% confidence limit of the estimated parameter.}
    \label{Fig:massmass}
\end{figure*}

\subsection{Rate of advance of periastron: $\dot{\omega}$}
The very low significance $\dot{\omega}$~=~1.5(1.3)$\times10^{-3}$ deg yr$^{-1}$ measured with the T2 model, can still be used as an upper limit for the total mass of the system. 

When GR is assumed, the $\dot{\omega}$ can be described using the Keplerian parameters and total mass of the system as:
\begin{equation}
   \dot{\omega} = 3 T_\odot^{2/3}\frac{P_{\rm b}}{2 \pi}^{-5/3}\frac{1}{1-e^2}(M_{\rm p}+ M_{\rm c})^{2/3}.
\end{equation}
Using the masses obtained from the T2 model, we obtain a prediction of $\dot{\omega} = 2.0(3) \times 10^{-4}\, \rm \deg \, yr^{-1}$. The observed value is consistent with this expectation but is not yet precise enough to usefully constrain the masses (see Fig.~\ref{Fig:massmass}).

\subsection{Shapiro delay}

As seen in Fig. \ref{Fig:Residuals}, the Shapiro delay signal, which describes the relativistic light-propagation delay in the system, has a maximum of $\sim$~9~$\mu$s using the MeerKAT data. This low amplitude is a consequence of a more face-on configuration of the system, and one of the reasons why this delay was not detected
until now. Using the ELL1H model, we have found a significant detection of $h_3 = 0.307^{+0.022}_{-0.026}$ $\mu$s and an orthometric ratio $\varsigma = 0.551^{+0.057}_{-0.054}$.
These constraints are displayed in Fig.~\ref{Fig:massmass}.

\subsection{Change in projected semi-major axis: $\dot{x}$}
Using the ELL1H model, we measure a highly significant $\dot{x}~=~-202.1^{+2.5}_{-2.7}\times10^{-16}$ s s$^{-1}$. This measurement is fully consistent with the value presented by \citet{2023A&A...678A..48E} but is 3 times more precise, going from a $\sim$~33 sigma detection to a $\sim$~100 sigma detection; an improvement due to the inclusion of the MeerKAT data. This $\dot{x}$ can be the result of various effects and is summarised as follows:

\begin{equation}
    \left( \frac{\dot{x}}{x} \right)^{\rm obs} = \left( \frac{\dot{x}}{x} \right)^{\rm GW} + \left( \frac{\dot{x}}{x} \right)^{\rm pm} + \left( \frac{d\varepsilon_A}{dt} \right) - \left(\frac{\dot{D}}{D} \right) +\left( \frac{\dot{x}}{x} \right)^{\dot{m}}+ \left( \frac{\dot{x}}{x} \right)^{\rm SO} + \left( \frac{\dot{x}}{x} \right)^{\rm planet},
\end{equation}
where the contributions to the observed $\dot{x}$ are due to the emission of gravitational waves (GW), proper motion of the binary system ($\mu$), varying aberration, $d\varepsilon_A/dt$, changing Doppler shift $-\dot{D}/D$, spin–orbit coupling and a hypothetical third companion (planet) around the pulsar.

\begin{figure}
   \centering
   \includegraphics[width=0.5\textwidth]{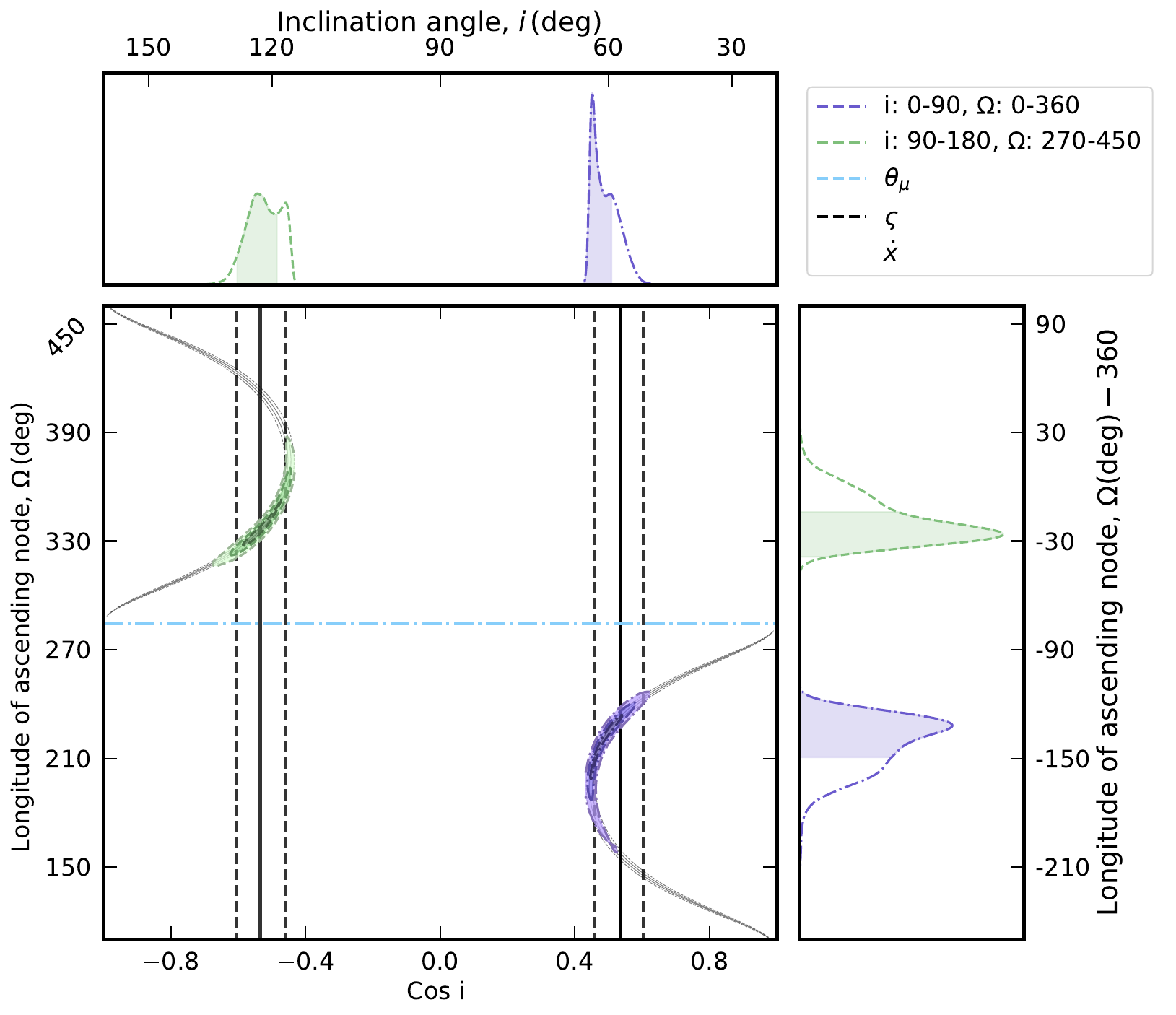}
      \caption{Plot of the allowable orbital orientation of J1455$-$3330. The black lines represent constraints from $\varsigma$ and gray contours show the constraint from $\dot{x}$ (1-$\sigma$ solid, 3-$\sigma$ dashed). The green dashed contours show the constraints from \texttt{temponest} using priors where the inclination angle ranges from 90 to 180 and $\Omega$ ranges from 270 to 450 (-90-90$^\circ$). The purple dashed contours show the constraints from \texttt{temponest} using priors where inclination angle, $i$, ranges from 0 to 90 and longitude of ascending node $\Omega$ ranges from 0 to 360. The light blue dashed horizontal line shows the position angle of the proper motion. The marginalised constraints on $\cos{i}$ and $\Omega$ are shown as 1D histograms (in green and purple) in the top and side panels.}
         \label{Fig:3doo}
   \end{figure}

A detailed calculation shows that only the secular variation of $x$ caused by the proper
motion ($\dot{x}/x)^{\rm pm}$, \citealt{1996ApJ...467L..93K}) contributes significantly to the observed $\dot{x}$. The term arises because our viewing angle of the binary ($i$) keeps changing due to its proper motion. The effect on $x = a{_p} \sin{i}/c$ is given by:
\begin{equation}
    \dot{x}_{\rm pm} = \mu \cot{i} \sin{(\Theta_{\mu}-\Omega)}
\end{equation}
The constraints derived from $\dot{x}$ on the orbital orientation of the system can be seen as dotted lines in Fig.~\ref{Fig:3doo}.

\subsection{Annual orbital parallax}
As the Earth orbits the Sun, a binary system is observed at slightly different angles, which can be seen as a periodic change in the apparent inclination angle (and thus of $x = a_{\rm p} \sin i$) of the pulsar orbit. For nearby binary systems, this effect is known
as the annual orbital parallax \citep{1995ApJ...439L...5K}, and can cause a measurable variation of the projected semi-major axis. Following \citet{1995ApJ...439L...5K}, this cyclic effect can be expressed as,
\begin{equation}
    \Delta_{\pi} = -\frac{1}{cd} \left(r \cdot r_{p} - (K_0r) (K_0r_p) \right),
\end{equation}
where c is the speed of light, d is the distance between the binary and the Solar System barycentre (SSB), and $K_0$ is the unit normal vector pointing from the SSB to the barycentre of the binary. The vectors $r$ = (X, Y, Z) and $r_{\rm p}$ are the Earth’s position with respect to the SSB and the pulsar position with respect to the SSB, which depend on the Solar System ephemeris model that is used and varies with time. Following \cite{2023A&A...674A.169G}, we estimate the expected peak-to-peak amplitude of the annual orbital parallax (AOP) by assuming that both the pulsar’s binary orbit and Earth’s orbit are circular ($e$ = 0). We use 
\begin{equation}
\begin{split}
\Delta_\pi = \frac{x_{\rm p}}{d_{\rm{PSR}}} \Bigl[
(\Delta_{\rm I0} \sin{\Omega} - \Delta_{\rm J0} \cos{\Omega}) \sin{\omega_{\rm{Pb}} t} \cot{i} \\
- (\Delta_{\rm I0} \cos{\Omega} + \Delta_{\rm J0} \sin{\Omega}) \cos{\omega_{\rm{Pb}} t} \csc{i} 
\Bigr]
\end{split}
\end{equation}

where x$_p$ is the projected semi-major axis, $\Omega$ is the longitude of the ascending node, $i$ is the inclination angle of the system, and $\omega_{\rm{Pb}}$ = 2$\pi$/P$_{\rm b}$ is the binary orbital frequency. This effect may be prominent in J1455$-$3330 since the projected semi-major axis, $x_{\rm p} = 32.3622059(4)$ s, is large, and the system is relatively close to the Earth ($\sim$ 0.90(5) kpc).

Measuring this effect will help break the degeneracy in the 3D orbital geometry of the system that results, in Fig~\ref{Fig:3doo} from the multiple intersections of the $\varsigma$ and $\dot{x}$ lines. The unit vectors ($I_0$, $J_0$, $K_0$) describe the coordinate system of the pulsar reference frame, with its origin at the binary system barycentre. Following \cite{1995ApJ...439L...5K},
\begin{align}
    \Delta_{I_{0}} &= (r \cdot I_0)  = -X \sin{\alpha} + Y \cos{\alpha}
\\    
    \Delta_{J_{0}} &= (r \cdot J_0)  = -X \sin{\delta}  \, \cos{\alpha} - Y \sin{\delta} \, \sin{\alpha},
\end{align}
where $\alpha$ and $\delta$ are the pulsar's right ascension and declination, and r = (X, Y, Z) is the Earth’s position with respect to the SSB described before. We obtain the Earth's (X, Y, Z) coordinates as a function of our observing MJD range using the same JPL solar ephemeris as in our timing results (DE438), which is contained within the \texttt{jplephem} package and implemented in \texttt{astropy}. We note that  \citet{1995ApJ...439L...5K} generally follows the coordinate system of DT92, and the relevant angles need to be transformed accordingly to agree with the observer's convention used throughout this paper. We compute the resulting $\Delta_\pi$ oscillatory trend as a function of MJD and find a peak-to-peak orbital parallax of $\varpi \sim400$ ns. Therefore, the AOP signal is undetectable above the timing noise.

\begin{figure}[htb]
   \centering
   \includegraphics[width=\columnwidth]{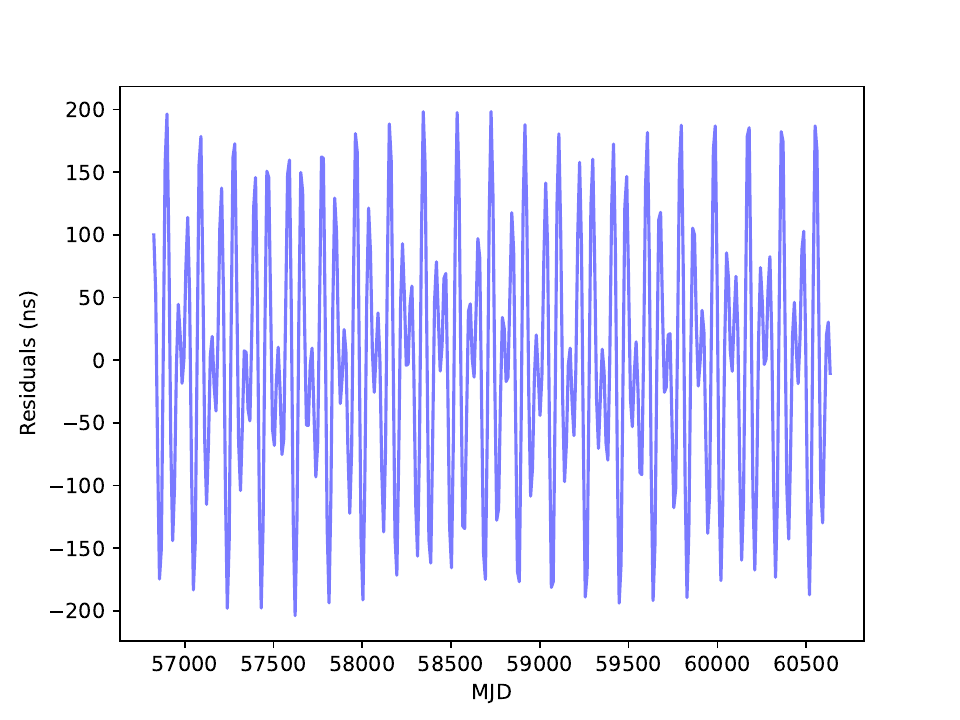}
      \caption{Simulated estimates of the peak-to-peak amplitude of the AOP for J1455$-$3330 ($\sim$ 400 ns). Our precision on $x_{\rm p}\sim0.4$~$\mu$s. Therefore, the AOP signal is undetectable above the timing noise.}
         \label{Fig:sims}
\end{figure}

\subsection{Self-consistent estimates of masses, and orbital orientation}\label{sec_5.9}
We now use all the effects we have been discussing until now in a self-consistent way to determine the masses of the pulsar, of the companion and the orbital orientation using \texttt{temponest}. The resulting constraints placed on $M_{\rm c}$, $\cos{i}$, and $M_{\rm p}$ are shown in Fig. \ref{Fig:massmass} as probability contours,
the constraints on $\cos{i}$ and $\Omega$ are shown in Fig.~\ref{Fig:3doo}.

The probability density contours for each of the islands show that two distinct regions in the $M_{\rm c}$-$M_{\rm p}$ plane and in the $\cos{i}$ - $\Omega$ plane are preferred. 
In all subplots of these Figures, the purple and green contours are given by the two separate \texttt{temponest} runs. TN run 1 has priors on $i$:0-90$^\circ$ and $\Omega$:0-360$^\circ$, and TN run 2 $i$:90-180$^\circ$ and $\Omega$:270-450$^\circ$. For the first run, we estimate $M_{\rm p} = 1.39^{+0.38}_{-0.18} \, \rm M_{\odot}$, $M_{\rm c} = 0.293^{+0.056}_{-0.026}\, \rm M_{\odot}$, $i = 63(2)^{\circ}$ and  $\Omega = 212(12)^{\circ}$.
For the second run we estimate $M_{\rm p}~=~1.53^{+1.10}_{-0.22} \, \rm~M_{\odot}$, $M_{\rm c}~=~0.309^{+0.163}_{-0.026}\, \rm M_{\odot}$, and $i~=~123(4)^{\circ}$ and $\Omega~=~334(12)^{\circ}$. 

In Fig. \ref{Fig:massmass}, we see how the masses allowed by the \texttt{temponest} runs follow the $h_{3}$ and $\varsigma$ constraints determined in the ELL1H solution. Although of low significance, the $\dot{\omega}$ constraints might in the near future set an upper limit on $M_{\rm p}$, excluding the large tail portion of the probability density contours.

Similarly, in Fig. \ref{Fig:3doo}, we see how the orbital orientations allowed by the \texttt{temponest} runs follow closely the $\dot{x}$ curve and $\varsigma$ constraints derived from the ELL1H model $\varsigma$. In this figure, we also see that the annual orbital parallax strongly disfavours values of $\Omega$ that are more than 90 degrees away from the proper motion.
At the moment, those are not taken into account given the very low statistical significance of $\dot{\omega}$.

In both analyses, the higher probability island includes the prior region $i$:0-90$^\circ$ and $\Omega$:0-360$^\circ$; the maximum likelihood of which are enclosed by purple contours in Fig. \ref{Fig:massmass} and Fig. \ref{Fig:3doo}.

\section{Discussion and conclusions} \label{Section6}
In this paper, we have presented the results of our timing analyses of J1455$-$3330 by combining available Lovell, NRT, Green Bank, and MeerKAT data with a total timing baseline of $\sim$30 years. The results include precise astrometry and parallax measurements, several kinematic effects, and a first measurement of its Shapiro delay.

A detailed analysis of the above relativistic effects has resulted in two solutions of the system's component masses and orbital orientation. Assuming general relativity we find two solutions: (1) a pulsar mass $M_{\mathrm p}~=~1.39^{+0.38}_{-0.18}\, \rm M_{\odot}$, a companion mass ${M_{\mathrm c}}~=~0.293^{+0.056}_{-0.026} \rm M_{\odot}$, an orbital inclination, $i$~=~63(2)$^{\circ}$, and longitude of the ascending node, $\Omega$~=~212(12)$^{\circ}$ or (2)  a pulsar mass $M_{\mathrm p}$~=~$1.53^{+1.10}_{-0.22}$ M$_{\odot}$, a companion mass ${M_{\mathrm c}}$~=~0.309$^{+0.163}_{-0.026}$ M$_{\odot}$, an orbital inclination, $i$~=~123(4)$^{\circ}$, and longitude of the ascending node, $\Omega$~=~334(12)$^{\circ}$. The mass measurements and orbital geometry of the system from both \texttt{temponest} runs are consistent. The best fit RVM model resulted in $\alpha=105.0(9)$, and $\zeta'=112.4(9)$, which translates to $i\sim180-\zeta'=61.6(9)$.

The companion mass in the higher probability island, $M_{\rm c}~=~0.293^{+0.056}_{-0.026}$ M$_{\odot}$, is consistent with a HeWD companion mass predicted by the \citet{1999A&A...350..928T} relation where the uncertainty on the measured companion mass ($\sim$0.046 M$_{\odot}$), gives a companion mass ranging from 0.247 M$_{\odot}$ to 0.339 M$_{\odot}$. In Fig. \ref{Fig:massmass}, the lower, middle and upper dashed light blue lines correspond to the predictions of $M_{\rm c}$ for Population I progenitors (corresponding to a metallicity of Z~=~0.02), Population II progenitors (Z~=~0.001), and Population I+II progenitors. For an orbital period of $\sim$ 76.17 days, the \citet{1999A&A...350..928T} relation predicts $M_{\rm c}\sim 0.315 \rm M_{\odot}$ given a Population I progenitor, $M_{\rm c}~=~0.348 \, \rm M_{\odot}$ given a Population II progenitor, and $M_{\rm c}~=~0.331\, \rm M_{\odot}$ given a Population I+II progenitor. A previous study regarding wide-orbit binary MSPs suggested that the \citet{1999A&A...350..928T} relation may overestimate the WD masses \citep{2005ApJ...632.1060S}. This study focused on binaries that did not contain precise mass measurements, and J1455$-$3330 is an example of a wide orbit binary whose estimated companion mass is in agreement with the \citet{1999A&A...350..928T} relation.

The measured pulsar mass in the higher probability island, $M_{\rm p}~=~1.39^{+0.38}_{-0.18}\, \rm M_\odot$, lies within 1$\sigma$ of the peak of the birth mass distribution proposed by \citet{2025NatAs.tmp...55Y}. \citet{2025NatAs.tmp...55Y} found that the neutron star birth mass function can be described as a turn-on power law model that has a minimum mass of $1.1^{+0.04}_{-0.05}\, \rm M_{\odot}$ and a maximum mass of $2.36^{+0.29}_{-0.17}\, \rm M_\odot$. The distribution peaks at  $1.27^{+0.03}_{-0.04}\, \rm M_\odot$ and declines with a power-law index of $6.5^{+1.3}_{-1.2}$. Electron-capture supernovae are likely to produce neutron stars with masses around the peak of the distribution, like J1455$-$3330. 

The eccentricity of 1.6$\times$10$^{-4}$ agrees well with the theoretical range predicted
by \citet{1994ARA&A..32..591P} for a MSP HeWD binary with an
orbital period of $\sim$76.17 days. The mass measurements, eccentricity, \citet{1999A&A...350..928T} relation and \citet{1994ARA&A..32..591P} relation, suggest that the system followed a standard evolutionary path, where the system is likely formed from wide-orbit low-mass X-ray binaries.

\begin{figure}[h]
   \centering
   \includegraphics[width=\columnwidth]{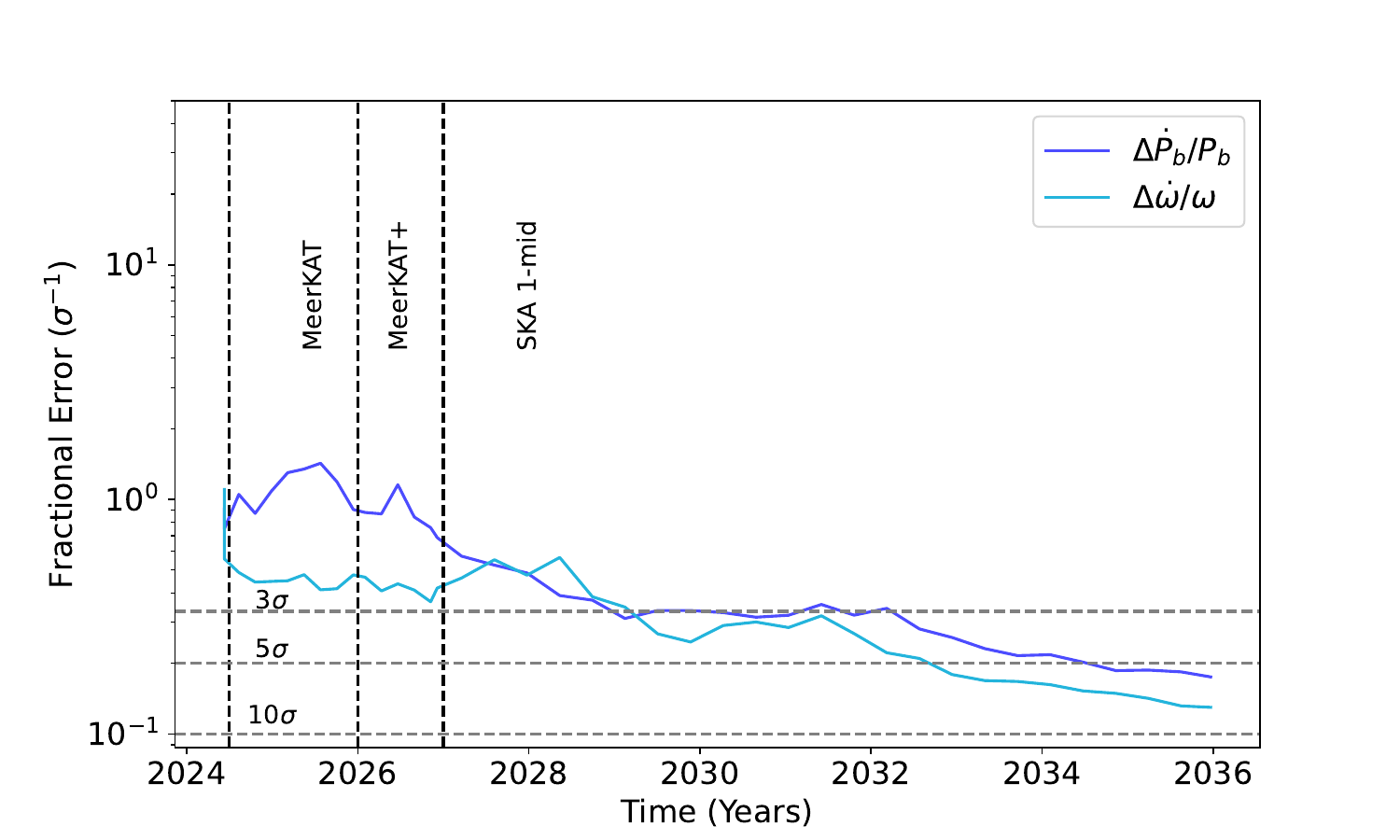}
      \caption{Improvement in the fractional error of orbital period decay, $\dot{P}_{\rm b}$, and $\dot{\omega}$ based on the simulations described in Sect. \ref{Section4}. The improvement in detection significance of the orbital period derivative ($\dot{P}_{\rm b}$) is shown in purple and the relative advance of periastron ($\dot{\omega}$) is shown in light blue. The vertical lines show the different simulated data; MeerKAT, MeerKAT+, and SKA 1-mid and the horizontal gray dashed lines show the significance thresholds.}
         \label{Fig:sims_improvement}
\end{figure}

The advance of periastron predicted by GR, $\dot{\omega_{\rm GR}}~=~2.0(3)~\times 10^{-4}$, and our fit for this parameter yields $\dot{\omega}~=~1.5(1.3)~\times10^{-3} \deg \, \rm yr^{-1}$. After $\sim$30 years of timing data, its uncertainty is still around two times larger than the expected effect (see Fig. \ref{Fig:sims_improvement}). However, measuring this parameter should be feasible in the not-too-distant future with continued timing with MeerKAT and SKA observations.

Indeed, simulated ToAs for J1455$-$3330 using knowledge of the sensitivity of current and future telescopes suggest a 3$\sigma$ detection of $\dot{\omega}$ around 2029 and a $3\sigma$ detection of $\dot{P}_{\rm b}$ around 2032. We assume future observing plans using the best telescopes for observing this pulsar, i.e. MeerKAT, MeerKAT+ and SKA 1-mid. We scale our ToA uncertainties for MeerKAT+ and SKA 1-mid based on our current ToA uncertainties measured with MeerKAT, and adopt an observing cadence of once every 7 days for MeerKAT and MeerKAT+, and once every 14 days for SKA 1-mid. A precise $\dot{\omega}$ measurement will result in a much more precise constraint on the component masses, and a measurement of $\dot{P}_{\rm b}$ will provide an independent estimate of the distance to the system.

Mass measurements of pulsars and companions in binary systems are important for the study of EoS, binary evolution theories and allows us to test key correlations such as the \citet{1999A&A...350..928T} relation.
The mass constraints of PSR J1455$-$3330, presented in this study, have provided another data point to the growing population of measured neutron star masses, the study of which will eventually help us identify the maximum possible mass of a neutron star and unveil more about the NS mass distribution itself. 

\begin{acknowledgements}
We thank the anonymous referee for the helpful comments and suggestions. The MeerKAT telescope is operated by the South African Radio Astronomy Observatory, which
is a facility of the National Research Foundation, an agency of the Department of Science and Innovation. SARAO acknowledges the ongoing advice and calibration of GPS systems by the National Metrology Institute of South Africa (NMISA) and the time space reference systems department of the Paris Observatory. MeerTime data is housed on the O-zSTAR supercomputer at
Swinburne University of Technology maintained by the Gravitational Wave Data Centre and ADACS via NCRIS support. Pulsar research at the Jodrell Bank Centre for Astrophysics and the observations using the Lovell Telescope are supported by a Consolidated Grant (ST/T000414/1) from the UK’s Science and Technology Facilities Council (STFC).

The Nan\c{c}ay Radio Observatory is operated by the Paris Observatory, associated with the French Centre National de la Recherche Scientifique (CNRS). We acknowledge financial support from the ``Action Th\'ematique de Cosmologie et Galaxies'' (ATCG), ``Action Th\'ematique Gravitation R\'ef\'erences Astronomie M\'etrologie'' (ATGRAM) and ``Action Th\'ematique Ph\'enomènes Extr\^emes et Multi-messagers'' (ATPEM) of CNRS/INSU, France. The Green Bank Observatory is a facility of the NSF operated under cooperative agreement by Associated Universities, Inc. 

This research has made extensive use of NASA's Astrophysics Data System (https://ui.adsabs.harvard.edu/). The analysis done in this publication made use of the open source pulsar analysis packages \texttt{psrchive} \cite{2004PASA...21..302H}, \texttt{tempo2} \cite{2006MNRAS.369..655H} and \texttt{temponest} \cite{2014MNRAS.437.3004L}, as well as open source Python libraries including Numpy, Matplotlib, Astropy and Chainconsumer. 

The authors thank T. Dolch, M. T. Lam, and E. Fonseca for their valuable comments, which helped improve this work. D.S.P. and V.V.K. acknowledge continuing valuable support from the Max Planck Society. D.S.P. acknowledges support from the International Max Planck Research School (IMPRS) for Astronomy and Astrophysics at the University of Bonn and University of Cologne. V.V.K. acknowledges financial support from the ERC starting grant ‘COMPACT’ (Understanding gravity using a COMprehensive search for fast-spinning Pulsars And CompacT binaries, grant agreement no. 101078094). J.S. acknowledges the support from the University of Cape Town Vice Chancellor’s Future Leaders 2030 Awards programme and the South African Research Chairs Initiative of the Department of Science and Technology and the National Research Foundation. R.M.S. acknowledges support Australian Research Council Future Fellowship FT190100155, and the Australian Research Council Centre of Excellence for Gravitational Wave Discovery (CE170100004 and CE230100016.) 
\end{acknowledgements}

\bibliographystyle{aa}
\bibliography{Biblio.bib}

\end{document}